\documentclass[12pt,preprint]{aastex}
\usepackage{natbib,graphics,graphicx}

\slugcomment{12 January 2009 revision; accepted by AJ}

\shorttitle{8-D Bayesian Quasar Selection}
\shortauthors{Richards et al.}

\begin{document}

\title{Eight-Dimensional Mid-Infrared/Optical Bayesian Quasar Selection}

\author{
Gordon T. Richards,\altaffilmark{1,2}
Rajesh P. Deo,\altaffilmark{1}
Mark Lacy,\altaffilmark{3}
Adam D. Myers,\altaffilmark{4}
Robert C. Nichol,\altaffilmark{5}
Nadia L. Zakamska,\altaffilmark{6}
Robert J. Brunner,\altaffilmark{4}
W. N. Brandt,\altaffilmark{7}
Alexander G. Gray,\altaffilmark{8}
John K. Parejko,\altaffilmark{1}
Andrew Ptak,\altaffilmark{9}
Donald P. Schneider,\altaffilmark{7}
Lisa J. Storrie-Lombardi,\altaffilmark{3}
and Alexander S. Szalay\altaffilmark{9}
}

\altaffiltext{1}{Department of Physics, Drexel University, 3141 Chestnut Street, Philadelphia, PA 19104.}
\altaffiltext{2}{Alfred P. Sloan Research Fellow.}
\altaffiltext{3}{Spitzer Science Center, Caltech, Mail Code 220-6, Pasadena, CA 91125.}
\altaffiltext{4}{Department of Astronomy, University of Illinois at Urbana-Champaign, 1002 West Green Street, Urbana, IL 61801-3080.}
\altaffiltext{5}{Institute of Cosmology and Gravitation, Mercantile House, Hampshire Terrace, University of Portsmouth, Portsmouth, PO1 2EG, UK.}
\altaffiltext{6}{John N.\ Bahcall Fellow, Institute for Advanced Study, Einstein Drive, Princeton, NJ 08540.}
\altaffiltext{7}{Department of Astronomy and Astrophysics, The Pennsylvania State University, 525 Davey Laboratory, University Park, PA 16802.}
\altaffiltext{8}{College of Computing, Georgia Institute of Technology, 266 Ferst Drive, Atlanta, GA 30332.}
\altaffiltext{9}{Department of Physics and Astronomy, The Johns Hopkins University, 3400 North Charles Street, Baltimore, MD 21218-2686.}

\begin{abstract}

We explore the multidimensional, multiwavelength selection of quasars
from mid-IR (MIR) plus optical data, specifically from {\em
  Spitzer}-IRAC and the Sloan Digital Sky Survey (SDSS).
Traditionally quasar selection relies on cuts in 2-D color space
despite the fact that most modern surveys (optical and infrared) are
done in more than 3 bandpasses.  In this paper we apply modern
statistical techniques to combined {\em Spitzer} MIR and SDSS optical
data, allowing up to 8-D color selection of quasars.  Using a Bayesian
selection method, we catalog 5546 quasar candidates to an $8.0\mu$m
depth of $56\mu$Jy over an area of $\sim24$ deg$^2$.  Roughly 70\% of
these candidates are not identified by applying the same Bayesian
algorithm to 4-color SDSS optical data alone.  The 8-D optical+MIR
selection on this data set recovers $97.7$\% of known type 1 quasars
in this area and greatly improves the effectiveness of identifying
$3.5<z<5$ quasars which are challenging to identify (without
considerable contamination) using MIR data alone.  We demonstrate
that, even using only the two shortest wavelength IRAC bandpasses (3.6
and 4.5$\mu$m),
it is possible to use our Bayesian techniques to select quasars with
97\% completeness and as little as 10\% contamination (as compared to
$\sim$60\% contamination using colors cuts alone).  We compute
photometric redshifts for our sample; comparison with known objects
suggests a photometric redshift accuracy of 93.6\% ($\Delta z\pm0.3$),
remaining roughly constant when the two reddest MIR bands are
excluded.  Despite the fact that our methods are designed to find type
1 (unobscured) quasars, as many as 1200 of the objects are type 2
(obscured) quasar candidates.  Coupling deep optical imaging data,
with deep mid-IR data 
could enable selection of quasars in significant numbers past the peak
of the quasar luminosity function (QLF) to at least $z\sim 4$.  Such a
sample would constrain the shape of the QLF both above and below the
break luminosity ($L^*_Q$) and enable quasar clustering studies over
the largest range of redshift and luminosity to date, yielding
significant gains in our understanding of the physics of quasars and
their contribution to galaxy evolution.


\end{abstract}


\section{Introduction}

Enormous progress has been made in the last decade on understanding
the nature of active galactic nuclei (AGNs) and their role in the
lifecycle of galaxies.  The co-evolutionary behaviour of black holes
and spheroids, as implied by the $M-\sigma$ relation
\citep[e.g.,][]{fm00,gbb+00,tgb+02}, suggests that most massive
galaxies hosted an AGN at some time.  Indeed, energy injection from
AGN through so-called ``feedback'' mechanisms
\citep[e.g.,][]{sr98,fab99,beg04,hhc+06} may be the linchpin
connecting the blue, star forming and massive red, dead elliptical
galaxies, whether through direct energy input
\citep[e.g.,][]{bbm+06,csw+06}, or as a somewhat more coincident
product of the main quenching mechanism (e.g., major mergers could
drive both AGN and quenching; \citealt{kh00,hhc+08}).

Although feedback provides a key clue as to how galaxies evolve, there
is a great deal of degeneracy in quenching prescriptions
\citep[e.g.,][]{cdd+06}.  A promising avenue is constraining feedback
models by examining the luminosity dependence to AGN clustering in
combination with the AGN luminosity function (e.g., \citealt{lhc+06}).
However, current large quasar surveys typically track only the peak of
the quasar luminosity function (QLF) at $z < 2$ and only probe the
brightest quasars at $z > 3$, for example the Sloan Digital Sky Survey
(SDSS; \citealt{yaa+00}) and the 2-Degree Field Quasar Survey (2QZ;
\citealt{csb+04}).  Consequently, quasar clustering measurements
\citep[e.g.,][]{pn06,mbn+07,dsc+08} detect little luminosity
dependence at $z < 2$, and provide incomplete constraints at $z > 3$.

If we wish to have a complete picture of the true role of feedback in
quenching star formation and black hole growth in galaxies, then an
important tool is to probe AGN (in statistically significant numbers)
to luminosities below $L^*_Q$ at high redshift and below host galaxy
depths at low redshift.  In the long-term, the next generation of
survey facilities such as LSST \citep[e.g.,][]{lsst08}, Pan-STARRS
\citep{kab+02}, DES \citep{des05}, and VST/VISTA \citep{anp+07},
should produce sufficiently deep optical and near-IR photometry with
which to address this goal.  In the near-term, any sufficiently
large-area survey of AGN that probes a wider luminosity range will
blaze important observational and theoretical trails.  This is
particularly true of a large-area survey that contains the necessary
color baseline (typically mid-IR through optical/UV) with which to
simultaneously study AGN host galaxies at $z\lesssim1$ and to
characterize quasar photometric redshifts
\citep[e.g.,][]{rws+01,rbo+08} out to $z\sim3$ and beyond.  As such,
herein we describe a novel method for efficient selection of AGN from
the combination of optical and mid-IR data that allows one to probe to
the depths (and areas) that are required to compile the sort of data
set that we have highlighted.  As the existing overlap between SDSS
and mid-IR imaging 
grows, so too will our ability to collect statistically significant
samples of faint high-$z$ quasars that are needed to probe the
influence of quasar feedback on galaxy evolution.


In principle, differentiating quasars from stars using mid-IR (MIR)
colors is straightforward.  At MIR wavelengths, stars (excepting the
coolest brown dwarfs) are well-described by the Rayleigh-Jeans portion
of a blackbody spectrum ($f_{\nu}\propto\nu^2$), resulting in quite
blue colors, whereas quasars are much redder (steeper/softer;
$f_{\nu}\propto\nu^{0\; {\rm to}\; -2}$).  For MIR colors from {\em
  Spitzer} \citep{wrl+04}, it is not, in practice, quite this simple
because {\em Spitzer}'s (comparatively) large pixels ($1\farcs2$)
makes it difficult to distinguish point from extended sources, and
thus a simple color cut to select AGN suffers from considerable
contamination from quiescent galaxies.  However, judicious use of
additional color cuts such as applied by \citet{lss+04} and
\citet{seg+04} can be used to select quasars (both obscured and
unobscured) with good efficiency and completeness for relatively
bright MIR sources.  See \citealt{drp+08} for a recent review of AGN
selection techniques in the MIR.

In this paper we explore ways to improve upon color selection of AGNs
using MIR colors.  We particularly concentrate on 1) using
sophisticated Bayesian selection methods, to probe to fainter limits
(without additional contamination) than allowed by standard color
selection, 2) incorporating optical morphology information, and 3)
combining MIR and optical photometry, performing up to 8-D color
selection.  The latter approach increases completeness to
high-redshift quasars (particularly $3.5<z<5.0$) where standard MIR
color cuts are incomplete (or heavily contaminated).  In addition, we
will show that coupling deep optical and MIR data can overcome the
loss of the two longest wavelength IRAC bandpasses when {\em
  Spitzer's} coolant is depleted.  Indeed, selection of quasars (both
type 1 and type 2) and photometric redshift estimation for type 1
quasars suffer relatively little from this loss.  Future coupling with
deep UV, near-IR, and far-IR data from {\em GALEX}, UKIDSS, {\em
  Akari}, {\em WISE}, and {\em Herschel} will allow for further
improvements in both selection and photometric redshift estimation.

We structure the paper as follows.  \S~2 describes our sources of
data.  We review our Bayesian selection algorithm in
\S~\ref{sec:overview}, where we also describe the training sets
designed for this selection.  Application of our algorithm to combined
SDSS and {\em Spitzer} data sets is discussed in \S~\ref{sec:bayes}.
The resulting catalog is presented in \S~\ref{sec:cat}.  Obscured
quasars are discussed in \S~\ref{sec:type2} and we present our
conclusions in \S~\ref{sec:conclusions}.  Throughout this paper we
report photometry either in flux density (in Jy) or AB magnitudes
(denoted by square brackets).  For the latter, {\em Spitzer}-IRAC
Channels 1-4 are given by $[3.6]$, $[4.5]$, $[5.8]$ and $[8.0]$, which
are the nominal wavelengths of the bandpasses in microns.  The
conversion between AB and Vega ($[{\rm Vega}]-[AB]$) is taken to be to
be 2.779, 3.264, 3.748 and 4.382 mag for the four IRAC bandpasses.
For example $[3.6]-[4.5] ({\rm Vega}) = [3.6]-[4.5]({\rm AB}) +
0.485$.  Cosmology-dependent parameters are computed assuming
$H_o=70\,{\rm km\,s^{-1}\,Mpc^{-1}}$, $\Omega_m=0.3$ and
$\Omega_{\Lambda}=0.7$, in general agreement with the most recent WMAP
results \citep{dkn+08}.  Unless otherwise specified, the term
``quasar'' will refer to type 1 (broad-line) quasars/AGNs, regardless
of their luminosity.

\section{The Data}

\subsection{Samples}
The selection methods described in this paper will make use of a
variety of data sets.  The MIR data are drawn from publicly available
catalogs/images from all of the largest area {\em Spitzer} surveys
using the Infrared Array Camera (IRAC; \citealt{fha+04}) where there
exists overlap with optical data from the SDSS.  
We include {\em Spitzer}-IRAC
data from the XFLS \citep{lacy05}; SWIRE \citep{lsr+03}, specifically
the ELAIS-N1, ELAIS-N2, and Lockman Hole fields; the NOAO-Bo\"{o}tes
area from the IRAC Shallow Survey \citep{esb+04}; and S-COSMOS
\citep{scosmos07}.  The properties of these fields are summarized in
Table~\ref{tab:tab0}.
It is important to note that these {\em Spitzer}-IRAC data sets have
very different depths, as indicated in Table~\ref{tab:tab0}.
As the SWIRE data represent the largest fraction of objects in our
analysis, we adopt their 95\% completeness limit of 56$\mu$Jy at
8.0$\mu$m as the cutoff for our analysis.  This choice excludes some
high signal-to-noise data from the COSMOS field and keeps lower
signal-to-noise data from the XFLS field, but is a good compromise for
common analysis of all the data sets.  The specific IRAC catalogs used
are: XFLS (main\_4band.cat; \citealt{lacy05}), SWIRE ELAIS-N1, -N2,
and Lockman Hole (SWIRE2\_N1\_cat\_IRAC24\_16jun05.tbl,
SWIRE2\_N2\_cat\_IRAC24\_16jun05.tbl,
SWIRE2\_Lockman\_cat\_IRAC24\_10Nov05.tbl; Surace et al.\ 2005), and
COSMOS (COSMOS\_IRAC\_0407\_IRSA.tbl; \citealt{scosmos07}).  At the
time of writing there were no publicly available MIR source catalogs
for the NOAO-Bo\"otes data\footnote{Now available at
  http://data.spitzer.caltech.edu/popular/sdwfs/20081022\_enhanced/documentation/SDWFS\_DR1.html},
so we extracted photometry from the publicly available images when
constructing the quasar training set; see \S~\ref{sec:qsotrain}.  In
all there exists $\sim32$ deg$^2$ of overlap between wide-area {\em
  Spitzer} fields and SDSS, with another $\sim30$ deg$^2$ of SWIRE
data lacking SDSS coverage.

Our primary goal herein is to perform
8-dimensional\footnote{Technically, 8-color selection, but we prefer
  the term dimension over color so as to be clear our method is
  flexible enough to include information other than colors.} selection
of quasars (primarily type 1) using optical and MIR data sets.  The
eight dimensions refer to the 8 unique colors afforded by SDSS $ugriz$
magnitudes \citep{fig+96} and {\em Spitzer}-IRAC Channel 1-4 flux
densities.  In addition, we perform 6-D selection of quasars using all
5 SDSS filters and the 2 short wavelength IRAC bandpasses (since the
long wavelength bandpasses will not be available during {\em
  Spitzer's} warm mission after its coolant is exhausted;
\citealt{warm}).  As such, we compile two combined MIR+optical data
sets.  One includes all objects detected in both IRAC Channels 1 and 2
that are matched to SDSS sources (95\% completeness at $g=22.2$,
$i=21.3$), for 6-D color classification.  The second is a subset of
the first where the objects are additionally detected in both IRAC
Channels 3 and 4 (for 8-D color classification).  IRAC upper limits
are not considered as our selection algorithm is currently not
equipped to handle them.  Note, however, that the SDSS's use of asinh
magnitudes \citep{lgi+01} means that any object detected in one SDSS
bandpass will have meaningful magnitude measurements in all the other
bandpasses (even if they are nominally below the ``flux limit'').
Unless otherwise specified, all SDSS magnitudes herein are PSF
magnitudes that have been corrected for Galactic extinction according
to \citet{sfd98}.



\subsection{SDSS-{\em Spitzer} Matching}

\begin{figure}
\plotone{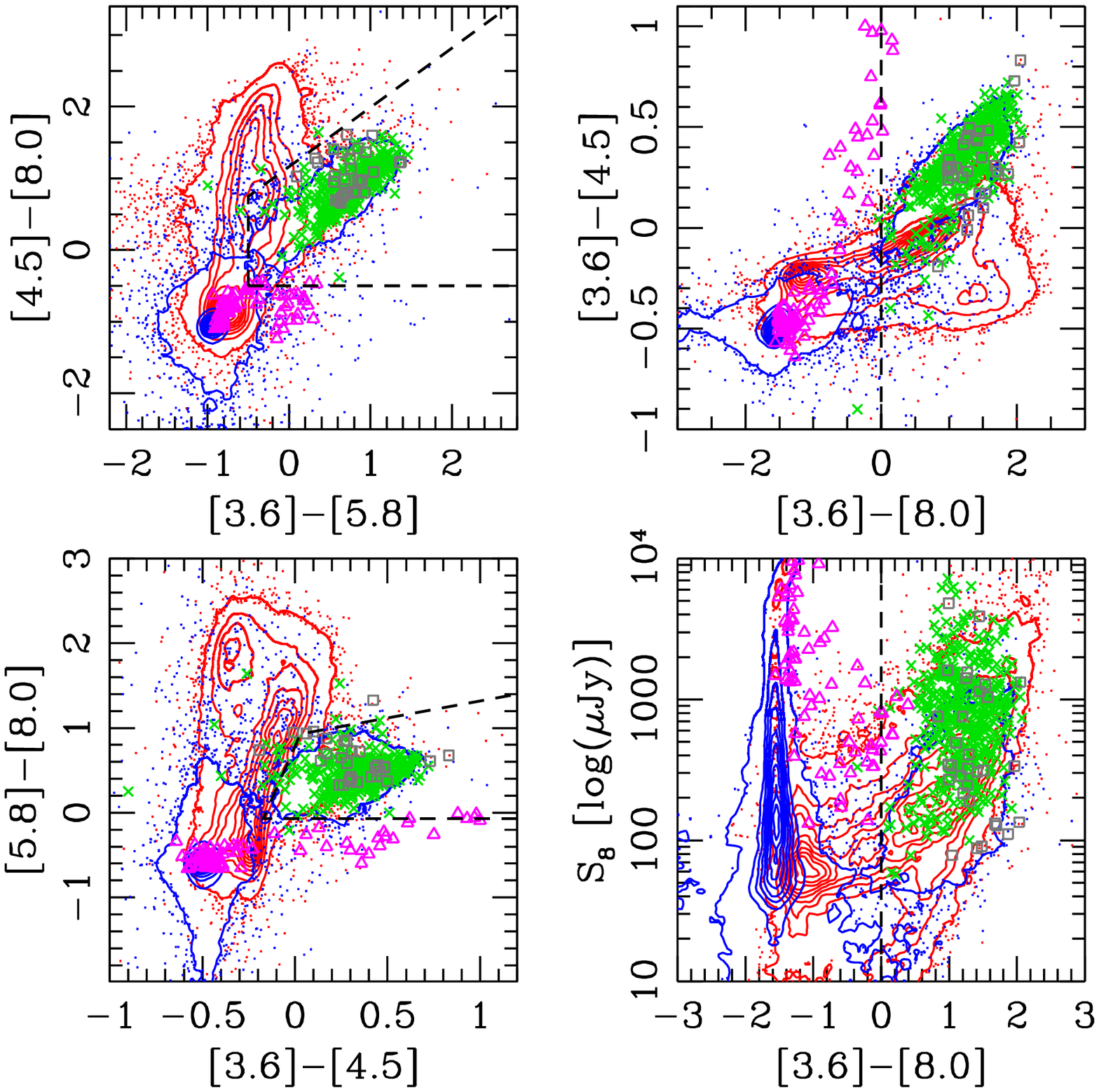} 
\caption{Comparison of MIR colors of point ({\em blue} contours/dots)
  and extended ({\em red} contours/dots) for various MIR color and
  flux combinations.  Green points depicts known type 1 quasars, while
  open grey squares are type 2 quasars.  Open magenta triangles
  indicate brown dwarfs \citep{psb+06}.  Comparison of these panels
  with Table~1 demonstrates that point sources with red MIR colors are
  robust AGNs candidates. 
The dashed lines depict the Lacy wedge region (upper left),
  Stern wedge region (lower left) and a simple $[3.6]-[8.0]$ color+morphology
  cut (right panels); see \S~\ref{sec:overview}.  Statistical errors are generally less than 1\%,
  but systematic errors can be $\sim10$\%.
\label{fig:iracstargal}}
\end{figure}

Matching the combined 2-band IRAC data to the SDSS optical imaging
catalog from the 6th SDSS data release (DR6; \citealt{sdssdr6}) with a
$2\arcsec$ matching radius (IRAC has $1\farcs2$ pixels and the median
SDSS seeing is $\sim1\farcs3$) yields 324,618 objects.  Of these, 486
objects are duplicates (243 pairs) where more than one SDSS source
matched an IRAC source.  To avoid any contamination, both objects in
all of these duplicates were rejected.  Spot checking of a few
duplicates revealed that they tended to be galaxies that were
improperly deblended in the SDSS.  The 324,132 matches obtained after
rejecting duplicates compose ``Sample A''.  Further limiting Sample A
to objects detected in all 4 IRAC bands leaves 95,181 objects; we
refer to this sample hereafter as ``Sample B''.  Sample B will be used
for our 8-D classifications (and in the construction of our training
sets).  The MIR colors of Sample B are shown in
Figure~\ref{fig:iracstargal}.

In Figure~\ref{fig:iracstargal}, the red contours/dots denote objects
classified as extended sources, while the blue contours/dots are point
sources. The extended vs.\ point-like classification is obtained on
the basis of SDSS optical data by comparing PSF magnitudes with measures of
extended flux for all bands in which the object is detected
\citep{slb+02}.  For faint optical sources, star-galaxy separation
begins to break down and galaxies become significant ``stellar''
contaminants.  At $i\sim20.8 \;(S_{8\mu \rm m}\sim115\mu {\rm Jy}$ at
$z=1.5$ for a type 1 quasars), roughly 10\% of SDSS point sources are
likely to actually be galaxies.  

The colors of the extended sources are concentrated in three clumps in
Figure~\ref{fig:iracstargal} (red contours/dots); understanding their
origins is important for optimal object classification.  The spectral
energy distribution of extragalactic sources in the 1-8$\mu$m range is
composed of three main components: (1) The combined light of stellar
photospheres has a peak at 1.6$\mu$m \citep{saw02} and declines
according to the Rayleigh-Jeans law at longer wavelengths; (2)
star-formation activity in the galaxy results in polycyclic aromatic
hydrocarbon (PAH) emission from dust with strongest features at 3.3,
6.2, 7.7 and 8.6$\mu$m; and (3) circumnuclear dust may be heated by
the central AGN resulting in continuum emission at MIR wavelengths,
which is typically well-represented by a power-law since the dust is
emitting at a wide range of temperatures.  The relative contributions
of these components to the total spectrum determines the IRAC colors
of extragalactic objects\footnote{MIR colors can also be affected by
  the 10$\mu$m silicate absorption/emission feature
  \citep[e.g.,][]{hss+05} at $z\sim0$ if the feature is very broad
  (IRAC channel 4 cuts off at $\sim9.5\mu$m).}.  The colors of
star-dominated and PAH-dominated galaxies at a wide range of redshifts
(0--1.6) are quite distinct in the IRAC bands from the colors of the
thermal emission of circumnuclear dust allowing color separation of
AGNs from galaxies of all types.  See \citet{bba+06} and
\citet{drp+08} for more details regarding how galaxies track through
MIR color space with redshift.

\subsection{Quasars}
\label{sec:qsotrain}

In addition to matching the IRAC catalogs to the full SDSS optical
catalogs, we have also explicitly matched it to the 77,429
spectroscopically confirmed quasars from the SDSS-DR5 quasar catalog
\citet{shr+07}.  As the density of these known quasars is much smaller
than the full SDSS object catalog, it is not too cumbersome to extract
IRAC photometry from the publicly available IRAC images of Bo\"{o}tes
field.  While our Bo\"{o}tes data reduction was simplistic compared to
the XFLS, SWIRE, and COSMOS data reductions, our analysis is
substantiated by the lack of any difference in color as a function of
redshift for the quasars in the Bo\"{o}tes region.

In all, matching the SDSS-DR5 quasar catalog to the IRAC data sets
(again using a $2\arcsec$ matching radius) resulted in 425 matches.
The relative IRAC and SDSS limits are such that all SDSS quasars are
detected in the MIR.  The additional 166 quasars as compared with
\citet{rls+06} come from the COSMOS and Bo\"{o}tes fields.  We
supplement these quasars with additional spectroscopically-confirmed
quasars with both {\em Spitzer}-IRAC and SDSS photometry from
\citet{lss+04}, \citet{pap+05}, and \citet{jfh+06}, where the latter
objects are $z\sim6$ quasars and the two former samples were
restricted to broad-line, type 1 AGNs.  In all we have compiled 515
known type 1 quasars with both SDSS and IRAC detections; the MIR
colors of these quasars are shown in Figure~\ref{fig:iracstargal} as
green points.

The SDSS-DR5 quasar catalog covers nearly all of the areas of overlap
between SDSS and the SWIRE fields. Therefore, with the exception of
several small-area fields and a part of the SWIRE/ELAIS-N1 area, these
objects represent essentially all of the spectroscopically confirmed
quasars that have been covered by both IRAC data and the final SDSS
data release (DR7).  The next major opportunity to obtain a large
sample of spectroscopically confirmed quasars with MIR follow-up
observations is with {\em WISE} \citep{WISE}, scheduled to launch in
late 2009.  {\em WISE}'s 120$\mu$Jy depth at 3.3$\mu$m will be
sufficiently sensitive to detect all of the $i<19.1$ quasars in the
SDSS catalog (but generally not the $z>3$ objects that extend as faint
as $i=20.2$) and will significantly increase the number of quasars
with MIR flux density measurements.

In addition to type 1 quasars, our selection is potentially sensitive
to obscured (type 2) quasars.  As such, it is important to know
where type 2 quasars lie in the MIR+optical color space that we
explore herein, so we have compiled a sample of type 2 quasars from
literature with measured redshifts (either photometric or
spectroscopic).  These include samples from \citet[][291
  objects]{zsk+03}; \citet[][887 objects]{rzs+08}; \citet[][6 objects
  from XFLS]{lsp+07}; \citet[][type 2 objects from Table 3, 30 objects
  from XFLS and SWIRE]{lps+07}; \citet[][one object at
  $z=3.7$]{nhg+02}; \citet[][20 objects from their Table 2, type 2 and
  unidentified objects (their type 9) with $\log(N_H) >
  21.8$]{mbh+02}; \citet[][one object at $z=3.288$]{smc+02};
\citet[][7 objects]{shl+06}; \citet[][125 objects]{pws+06};
\citet[][104 XMM-DS objects classified as AGN2 and starburst/AGN]{tpc+07};
\citet[][21 objects from SWIRE/NDWFS/FLS, their Table~1 with redshifts
  from their Table~3]{ptm+07}; and \citet[][141 objects from
  CDFS]{zmm+04}.  We then matched this combined sample with all the
available IRAC catalogs (including and in addition to those above)
to extract {\em Spitzer}-IRAC photometry.  Only objects detected at
both 3.6$\mu$m and 4.5$\mu$m were selected in the matched output.
This final output table was then matched to the SDSS DR6 catalog to
select objects with both SDSS and {\em Spitzer}-IRAC photometry.  This
filtering/matching process was used to generate a final list of 43
type 2 quasars/candidates from literature with both SDSS and {\em
  Spitzer}-IRAC photometry.  The MIR colors these type 2 quasars are
given by open gray squares in Figure~\ref{fig:iracstargal}.

\section{Bayesian Selection Overview}
\label{sec:overview}

Our selection method follows that of \citet{rng+04} and is detailed
therein. For multi-band imaging surveys (such as the SDSS) it is
possible to perform object classification beyond traditionally adopted
2-D color cuts.  In particular, if the parameter space of interest is
sufficiently populated by known objects, then these objects can be
used as a ``training set'' to derive classifications within the
parameter space \citep[e.g.,][]{htf01,rng+04,bbm+06,gzz08}.  Our
algorithm of choice is based on kernel density estimation (KDE),
weighted by a Bayesian ``prior".  The KDE aspect is that the
probability distribution function (pdf) that is used to evaluate the
classification is smoothed by some appropriate kernel function (e.g., a
Gaussian).  Our algorithm uses two training sets, one that represents
the objects to be classified and one that represents everything else.
We compute the N-dimensional Euclidean color distance between some new
object that we wish to classify and each of the training set
objects. The new object is then classified based on how consistent its
colors are with each training set \citep[e.g.,][]{grn+05}.  One
limitation of our current algorithm is that error information is not
explicitly included (cf., \citealt{pmh+07}), although it is {\em
  implicitly} included by virtue of the training sets' inherent error
distribution.

Our Bayesian selection algorithm is guided by four considerations.
First are the so-called ``wedge'' diagrams for MIR selection of AGNs.
These are regions of color space that tend to be occupied by AGNs.
\citet{lss+04} select AGNs using color cuts in non-adjacent bandpasses
to isolate AGNs, specifically $[3.6]-[5.8]$ and $[4.5]-[8.0]$; we
refer to these color cuts as the ``Lacy wedge''.
\citet{seg+04,ska+07} instead utilize adjacent bandpasses,
specifically $[3.6]-[4.5]$ and $[5.8]-[8.0]$; we refer to these color
cuts as the ``Stern wedge''.  The exact color-cuts that describe these
regions have changed slightly over time as more {\em Spitzer} data
have been obtained.  Thus we will refer to them in the abstract sense
throughout the paper, but we visually illustrate them in
Figure~\ref{fig:iracstargal}.  The Lacy wedge region is given by the
dashed lines in the top left-hand panel and the Stern wedge region is
given by the dashed lines in the bottom left-hand panel.


The second diagnostic considers the mean colors for stars and quasars
in the IRAC photometric system; Table~\ref{tab:tab1} gives both the
observed and theoretical (power-law approximated) values.  For stars
the $[3.6]-[8.0]$ color is $\sim-1.7$, while for quasars it ranges
from $\sim0$ to $1.7$.  Thus, well-measured point sources can be
grouped into stars and quasars in a straightforward manner, based on
MIR colors alone.

\begin{figure}
\plotone{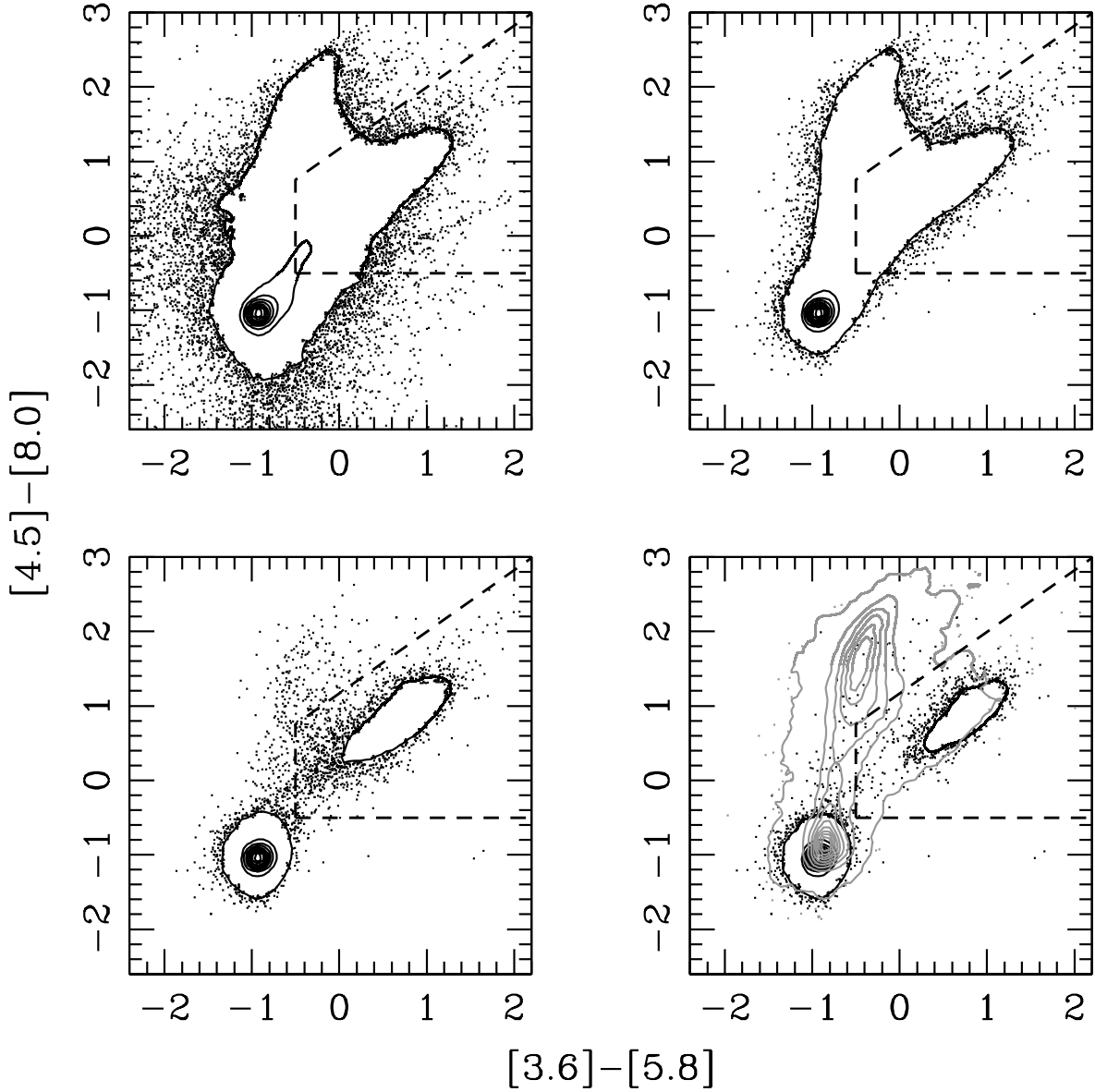}
\caption{Illustration of the power of combining {\em Spitzer} data
with morphology information.  Top left: All 4-band data.  Top-right:
Removed very faint and saturated objects.  Bottom-left: Limited to
SDSS point sources.  Bottom-right: Limited to SDSS point sources with
$g<21.5$ (grey contours indicate extended sources).  Dashed lines in
each panel show the Lacy wedge.  In short, MIR-only quasar selection is
robust only for relatively bright sources or when coupled with
accurate morphologies.  Similar results would be seen for the Stern
wedge (albeit with somewhat less contamination from galaxies at the
price of reduced completeness to high-$z$ quasars).
\label{fig:lacywedgemany}}
\end{figure}

Third, we consider the utility of having morphology information in
addition to photometry and the effects of over-reliance on morphology
as when star-galaxy separation fails to be robust.
Figure~\ref{fig:iracstargal} shows that AGNs are readily identified as
point sources with red MIR colors (even using only the two shortest
IRAC bandpasses).  However, photometric errors complicate clean AGN
identification at faint limits.  Using just the Lacy wedge as an
example, we demonstrate in Figure~\ref{fig:lacywedgemany} how using a
brighter MIR flux limit or morphology information can improve the
efficiency of MIR color selection.
Removing faint objects, saturated objects, and/or extended objects
significantly reduces contamination.  

Finally, we consider the color-redshift distribution for known
quasars.  Six color combinations are shown in
Figure~\ref{fig:czplotnew} as a function of redshift for the known
quasar samples described above and for two template quasar spectra.
The Stern (dotted) and Lacy (dashed) wedge color cuts (see
Fig.~\ref{fig:iracstargal}) are included in order to show their
effects on completeness with redshift.  Cuts in $[3.6]-[4.5]$ can
introduce incompleteness to $3.5<z<5.0$ quasars.  The $z-[3.6]$ color
changes rapidly between redshift 0 and 2 due primarily to a minimum in
the SED, the so-called ``$1\mu$m inflection'' \citep[e.g.,][]{ewm+94}.

\subsection{Training Sets}
\label{sec:train}

The core of our quasar training set is the 515 optically-selected,
spectroscopically-confirmed quasars that have existing {\em
  Spitzer}-IRAC photometry as discussed in \S~\ref{sec:qsotrain};
this set is dominated by the 425 quasars from the SDSS-DR5 quasar
catalog.  Due to some photometric errors (bad deblending, etc.)
objects with $[3.6]-[4.5]<-0.7$ or $[5.8]-[8.0]<-0.5$ are rejected.
As our primary goal is efficient N-dimensional selection of type 1
(broad-line) quasars, we have not explicitly included obscured or
intrinsically weak AGNs in the training set, although doing so would
be a logical next step for future investigation.

Since the number of known quasars with optical and MIR photometry is
relatively small in comparison with the size of the sample that we
wish to classify, we further include IRAC 4-band matches to SDSS
sources that are highly likely to be quasars.  Below we describe two
such classes of objects that are included.  For these samples to be as
clean as possible, we require that the sources satisfy $56\mu{\rm
  Jy}<f_{8.0\mu{\rm m}}<10{\rm mJy}$, and $i>14.0$ to exclude
saturated objects and to limit the samples to the approximate
8.0$\mu$m 95\% completeness limit of the SWIRE data.

First, given the power of combining morphology with MIR colors shown
in Figure~\ref{fig:lacywedgemany}, we identify point sources with red
(AGN-like) MIR colors as quasar candidates that are sufficiently
robust to be included in the training set.  As SDSS star-galaxy
separation begins to break down at faint magnitudes ($r\sim21$;
\citealt{sjd+02}) with faint extended sources being more likely to be
classified as point-like, clean point source identification should be
restricted to $i\lesssim20.8$ (0.5 mag brighter than the nominal SDSS
$i$-band flux limit).  Thus, point sources with $i<20.8$ are classified
as quasars if they have $[3.6]-[8.0]>0$, see Table~1.  However, for
the XFLS data, the MIR photometric errors are large enough that a
simple cut in $[3.6]-[8.0]$ is insufficient to cleanly identify
quasars; for these objects we also require $[3.6]-[5.8]>-0.5$ and
$[4.5]-[8.0]>-0.5$ (akin to the Lacy wedge).  Finally true point
source quasars do not typically have MIR colors outside of
$-0.4<[5.8]-[8.0]<1.4$ and $[3.6]-[4.5]<-0.6$, so we exclude such
objects to further limit any contamination by misclassified galaxies
and by normal stars.

Second, in addition to the above red MIR point sources, we capitalize
on the work of \citet{lss+04} and \citet{seg+04} by including objects
that can be robustly identified as quasar candidates based on their
location in the MIR wedge diagrams.  While selection using these
wedges is relatively clean at bright limits, the colors of quasars may
change with flux (indeed Fig.~\ref{fig:iracstargal} suggests that at
fainter limits the quasar contribution weakens with respect to the
galaxy contribution, making the MIR colors bluer on average), thus
additional constraints are needed to exclude contaminants.  In
particular, we include in the quasar training set any {\em point
  sources} that are in any of 1) the Lacy wedge, 2) the Stern
wedge, or 3) a modified Stern wedge (which excludes a region near
the galaxy locus, see \S~\ref{sec:wedge} below).  We further include
{\em extended sources} that lie in {\em both} the Lacy wedge and our
modified Stern wedge.  In addition to relatively robust MIR point
source candidates, these criteria select point sources near the
quasar/galaxy boundary in $[3.6]-[4.5]$ (between the original and
modified Stern wedges), but reject extended sources in this same
``buffer zone'' where it may be difficult to distinguish true galactic
contaminants from host-dominated quasars.

For these two MIR-selected training set populations, we have not
utilized any optical magnitude or color information since optical
quasar selection is known to be incomplete in certain redshift ranges
(e.g., $z\sim2.7$, \citealt{rsf+06}), and the inclusion of these
MIR-identified sources is our attempt to mitigate this effect.
However, roughly half of these objects
are UV-excess sources that would be included using optical selection
alone.  Similar to the $z\sim2.7$ redshift incompleteness in the
optical,
MIR-only selection is incomplete in a different redshift range
($3.5<z<5$), thus the inclusion of optical-only selection objects in
the training set helps recover such objects in our higher dimensional
selection.  Having a training set comprised of both optical-only
(the above 515 known quasars) and MIR-only identified sources aids in
the creation of a photometric quasar sample that is as complete as
possible at all redshifts.  While no MIR-only or optical-only quasar
samples are fully complete at all redshifts, by combining the two and
then performing quasar selection simultaneously on both MIR and
optical colors, we overcome the limitations inherent to each method
separately.

Due to photometric errors (bad deblending, etc.), a handful of the
quasar training set objects are outliers from the quasar
color-redshift distribution.  We only retain objects that are in the
intersection of the following criteria: $[3.6]-[8.0]>-0.5$,
$[3.6]-[4.5]>-0.5$, $[4.5]-[5.8]>-0.7$, and $[5.8]-[8.0]>-0.5$.  In
summary, the final quasar training set is the combination of 1) known
quasars, 2) red MIR objects with point-like optical morphologies, 3)
point/extended sources identified in two distinct MIR color wedges,
and 4) excluding some photometric error induced outliers and some
extended sources on the border between known quasars and galaxies.

As was the case for our purely optical selection \citep{rng+04}, a
more difficult task is to define the non-quasar training set (here
both stars and normal galaxies) as we do not have a sufficiently large
(and representative) sample of spectroscopically confirmed objects.
While there are many thousands of SDSS spectra of galaxies and stars,
the galaxy sample extends only to $r=17.77$ and the star spectra cover
only specific regions of color parameter space.  Furthermore,
optical spectra may not reveal the AGN nature of a galaxy as well as
the MIR colors would.  Thus, we are essentially left with identifying
the leftovers from our quasar training set as the non-quasar training
set, with the exception of borderline objects that we exclude
from either training set.  Objects in the non-quasar training set
satisfy the following conditions.  First, they must not be in the
quasar training set.  Second, they are rejected if they lie in the Lacy
wedge and also either the original or modified Stern wedges.  The
remaining objects that meet our magnitude and flux limits constitute
the non-quasar training set.

Figure~\ref{fig:iraccolorstrainlib} shows the MIR color distribution
of the objects in the quasar and non-quasar training sets.  The SDSS
colors of the same objects are shown in
Figure~\ref{fig:sdsscolorstrainlib}.  The training set has 53332
objects; 5627 labeled as quasars and 47705 as non-quasars.  For our
Bayesian classification method, we need to estimate a ``prior''
that indicates the a priori probability of any given object in our
samples not being a quasar.  We adopt the fraction determined from the
training sets as a reasonable value, specifically 89\%.



\subsection{The Modified Stern Wedge Region}
\label{sec:wedge}

In Figure~\ref{fig:wedgecomp} we compare and contrast the two primary
MIR color selection techniques in use today; see also \citet{drp+08}.
By exploring the color space occupied by objects selected by the Stern
wedge, but not the Lacy wedge, and vice versa, we hope to create a
more robust quasar training set from MIR colors.  Examination of
Figure~\ref{fig:wedgecomp} reveals the following
\begin{itemize}
\item The Stern wedge is relatively clean (few objects are outside of the Lacy wedge).
\item The Lacy wedge is very clean to the 1\,mJy flux limit for which
  it was defined but is quite contaminated at fainter MIR fluxes (it
  contains many objects outside of the Stern wedge).
\item The Stern wedge's lack of contamination comes largely from its
  cut on $[3.6]-[4.5]$. Although very effective over a wide range of
  redshifts, this cut makes the Stern wedge incomplete to quasars with
  $3.5 < z < 5.0$ (see Fig.~\ref{fig:czplotnew}).
\end{itemize}
Furthermore, we note that the Stern wedge's blue cut in $[5.8]-[8.0]$ excludes
objects that are in the Lacy wedge.  Such objects are rejected
nominally to exclude high-redshift galaxies (see Fig.~1 in
\citealt{seg+04}); however, as these objects are found by
\citet{hjf+06} to be strong soft X-ray sources, the AGN population
likely crosses this dividing line.

Indeed, during a 2008 June observing run on the Mayall 4-m telescope
at Kitt Peak National Observatory, we were able to make observations
of 9 randomly chosen sources blueward of the Stern wedge's
$[5.8]-[8.0]$ cut.  Of the six sources redder than $[3.6]-[4.5]=-0.1$,
three are clearly quasars ($z=0.91, 1.61, 1.64$) and a 4th may also be
a quasar at low signal-to-noise; their MIR colors are shown in
Figure~\ref{fig:wedgecomp}.

These objects are all in the region of strong soft X-ray sources (and
thus likely quasars) as indicated by \citet{hjf+06} and have two
similar features.  All are flagged as color selected quasar candidates
\citep{rfn+02} in the SDSS database, but were too faint to be targeted
for spectroscopy.
They also have very similar appearances in the optical, being
extended, but centrally concentrated bluish-white objects.  When
coupled with MIR data, these characteristics may help to identify
additional quasars in the SDSS database that are nominally outside of
the Stern wedge, but that may nevertheless be AGNs.



Thus, as mentioned above, we implement a ``modified Stern wedge'' that
is adjusted from the original Stern wedge as follows.  We first make a
more conservative cut on the AGN/galaxy border in $[3.6]-[4.5]$ color.
Second, we allow objects bluer than $[5.8]-[8.0]<-0.07$ if they have
$[3.6]-[4.5]>-0.0626$ (the intersection of our modified $[3.6]-[4.5]$
and the original $[5.8]-[8.0]$ cuts).  This modified Stern wedge is
defined by the following set of color cuts (see
Fig.~\ref{fig:wedgecomp})
\begin{eqnarray}
(([5.8]-[8.0])>-0.07 \; \&\& \; ([3.6]-[4.5])>0.18*([5.8]-[8.0])-0.05 \; \&\& \nonumber \\
([3.6]-[4.5])>2.5*([5.8]-[8.0])-2.295) \;||\; \nonumber \\
([3.6]-[4.5]>-0.0626 \; \&\& \; (-0.5<[5.8]-[8.0]\le-0.07)).
\end{eqnarray}
No modifications have been made to the Lacy wedge because coupling it
with the Stern wedge already removes spurious sources.  We are not
suggesting that this modified wedge be used in place of the Stern
wedge for MIR AGN selection, rather we use it to minimizing
contamination from normal galaxies in our training sets, while
attempting to maximize completeness to quasars with $3.5<z<5$.

\section{Application of the Algorithm}
\label{sec:bayes}

Once the training sets are defined, we follow the techniques described
in \citet{rng+04} for utilizing these training sets to select objects
from a sample of data.  We shall perform this selection for two sets
of color space: 8-D (MIR+optical) to attempt a more complete type 1
quasar selection than can be accomplished with MIR-only or
optical-only selection; and 6-D, anticipating the operation of {\em
  Spitzer} post-cryogen, when only IRAC Channels 1 and 2 will be
operational.  We have additionally attempted a 3-D MIR-only Bayesian
selection.  Although this approach adds an extra color dimension that
is otherwise being wasted by the 2-D MIR wedge selection methods, we
find that it does not work significantly better (e.g., it is still
rather incomplete to $3.5<z<5.0$ quasars as is the Stern wedge) and do
not discuss it further here.

In addition to a {\em prior} (as discussed above) our algorithm also
uses a leave-one-out cross-validation process to determine the optimal
bandwidths (kernel smoothing parameter) for the AGN and non-AGN
samples, respectively.  These bandwidths are essentially the
resolution of the pdf, akin to the bin size for a histogram.  See
\citet{rng+04} for details, but, in brief, the algorithm examines a
range of bandwidths and chooses the one that maximizes our
completeness to known quasars while minimizing contamination.  The
adopted training set bandwidths notated as (star, quasar) were
$(0.2,0.11)$ and $(0.195,0.05)$ magnitudes, for the 8-D and 6-D
selection, respectively.  The selection is reasonably robust to small
($\sim0.05$ mag) changes from these values.

\subsection{8-D}

Quasar selection is essentially an algorithm that identifies outliers
from the stellar locus in the optical or the galaxy locus in the MIR.
Thus, one might expect that combining optical and MIR photometry
together will yield more robust quasar selection than optical or MIR
photometry alone as objects only need to be outliers in one dimension.
In particular, our desire is to combine the 5 SDSS and all 4 IRAC
bandpasses together to recover the $2.5<z<3.0$ quasars lost by optical
selection due to stellar locus contamination and the $3.5<z<5.0$
quasars lost by MIR selection due to galaxy locus contamination.

We have applied our Bayesian selection method to the 8 unique colors
afforded by SDSS plus IRAC photometry in Sample B.  As with the
training sets, we limit Sample B to $56\mu{\rm Jy}<f_{8.0\mu{\rm
    m}}<10{\rm mJy}$, and $i>14.0$, which reduces the sample from
95,181 objects to 52,659 objects.  In all, 5468 quasar candidates were
found.

The MIR and optical color distributions of these 8-D selected objects
are shown in Figures~\ref{fig:iraccolorstest79Dclass} and
\ref{fig:sdsscolorstest79Dclass}, respectively.  Here we have added
theoretical power-law colors to the MIR color-color plots, which
demonstrates that the most robust objects tend to have power-law
colors in the MIR \citep[e.g.,][]{drp+08}.
The primary differences
between these and 2-D MIR wedge selection are that 541 objects outside
of the Stern wedge are now selected and the left (blue) part of the
Lacy wedge is more represented.  The redshift completeness (to known
type 1 quasars) is given in the left-hand panel of
Figure~\ref{fig:zhisttrainall}.  As compared to MIR-only selection
using the Stern wedge (right-hand panel), it is clear that that 8-D
MIR+optical selection performs better over $3.5<z<5$.  Overall the
completeness to spectroscopically confirmed type 1 quasars is 97.7\%.
This high completeness suggests that it should be possible to attempt
to classify fainter sources and still maintain a reasonably high
completeness --- although clearly photometric errors limit how
faint the method can be applied.



\subsection{6-D}

These results from our 8-D Bayesian selection are promising; however,
once {\em Spitzer} has exhausted its coolant, it will no longer be
able to observe in the two longest IRAC bands.  Thus an interesting
question is how well our methods will work when only IRAC Channels 1
and 2 are operational.

As such, we have also performed a 6-D Bayesian selection using the 5
SDSS bandpasses and only the 2 shortest wavelength IRAC bandpasses.
Instead of using Sample A here (which includes object rejected from
Sample B due to lack of Channel 3/4 detections), we instead have only
considered the same 52,659 magnitude/flux-limited sources from Sample
B above, simply ignoring the color information from Channels 3/4.  We
have made this choice in order to provide the most direct and unbiased
comparison between 8-D and 6-D selection.  However, we emphasize that,
similarly limiting sample A to unsaturated sources and the SWIRE 95\%
completeness limit at 3.6$\mu$m (Table~\ref{tab:tab0}) [allowing
  non-detections in Channels 3/4], we can potentially apply our 6-D
algorithm to over 290,000 sources.

Overall our 6-D algorithm selects 5222 objects and is not
significantly worse than 8-D, with a completeness of 97.1\% with
respect to the type 1 quasars in the training set.  The completeness
as a function of redshift (red line in the left-hand panel of
Fig.~\ref{fig:zhisttrainall}) is consistent with that for 8-D
selection.  The drop in $z>5.5$ completeness is not statistically
significant given the small number of objects considered.  In all, 324
(6\%) objects that were selected by the 8-D algorithm are not 6-D
candidates.  There are also 78 (1\%) 6-D selected sources that are not
8-D candidates.  Figures~\ref{fig:iraccolorstest79Dclass} and
\ref{fig:sdsscolorstest79Dclass} show the locations of objects
selected by both the 8-D and 6-D algorithms and also those selected by
only one.

\subsection{Comparison of Selection Methods}

Table~\ref{tab:tab2} shows a comparison between our Bayesian selection
methods, the Lacy and Stern wedges, and a simple $[3.6]-[4.5]>-0.1$
color cut.  For bright flux limits ($f_{8.0\mu{\rm m}}>1\,{\rm mJy}$)
our Bayesian method agrees well with the Lacy and Stern wedges, with
the $[3.6]-[4.5]>-0.1$ cut being about 50\% contaminated.  At
fainter limits the Lacy wedge is seen to be contaminated by inactive
galaxies, the $[3.6]-[4.5]$ cut somewhat less so.  Of 8-D Bayesian
selected quasars candidates, 99\% are within the Lacy wedge and 90\%
are within the Stern wedge; thus we expect the Bayesian sample to be
quite clean.

In terms of completeness, Figure~\ref{fig:zhisttrainall} demonstrates
that the Lacy wedge and either of our 8-D or 6-D Bayesian algorithms
are quite complete to type 1 quasars at nearly all redshifts.  The
comparison sample is the set of spectroscopically-confirmed type 1
quasars that were used to construct our quasar training set.  Thus,
Figure~\ref{fig:zhisttrainall} represents the self-selection
completeness.  As an additional check, we find that our algorithm
recovers all 51 of the type 1 quasars cataloged by \citet{tim+07} in
the COSMOS field (to $i<21.3$ and $f_{8.0\mu{\rm m}}<56\,\mu$Jy).  The
Lacy wedge is much more contaminated (Table~\ref{tab:tab2}) than our
Bayesian-selected samples.  The Stern wedge would appear to be the
least contaminated, but is also the least complete, particularly for
$3.5<z<5.0$.  A simple $[3.6]-[4.5]$ color cut is somewhat more
complete than Stern wedge selection, but suffers from considerable
contamination.  While a $[3.6]-[8.0]$ color cut might be expected to
be cleaner than $[3.6]-[4.5]$, the presence of PAH features at $8\mu$m
causes contamination from low-redshift PAH-dominated galaxies that
overwhelms the loss of stellar contaminants due to the longer
baseline.

A particularly interesting question is how well our MIR+optical
selection performs above optical-only selection algorithm.  For this
comparison, we utilize our optical-only Bayesian-selected quasar
catalog \citep{rmg+09}, which includes unresolved SDSS quasar
candidates to the nominal SDSS magnitude limit of $i<21.3$.  We find
that 1702 of 2426 (70\%) of our optical+MIR selected targets are also
selected by the optical-only algorithm (for $i<21.3$ point sources).
However, our MIR+optical selection benefits by adding 2117 extended
sources and 1003 points sources with $i>21.3$ (i.e., nominally fainter
than the SDSS magnitude limit).  Thus, using the same SDSS dataset as
a basis, our MIR+optical selection improves quasar selection by a
factor of 3.25 (5546/1702).

It is instructive to consider the future of MIR-based selection of
quasars in general, even without the benefit of our Bayesian selection
method.  Indeed, while the wedge selection methods rely on IRAC
Channels 3 and 4, it is actually Channels 1 and 2 that are most needed
--- in terms of separating AGNs from stars and normal galaxies.  As
can be seen in Figure~\ref{fig:iracstargal}, in the absence of
accurate morphological classification, quasars selected with a
$[3.6]-[4.5]$ cut will have less contamination from galaxies than a
$[3.6]-[8.0]$ cut, yet be similarly effective in removing stars.
Using a $[3.6]-[4.5]$ color cut alone, we find that for
$[3.6]-[4.5]>0.1$, 91\% of known quasars are recovered.  Relaxing the
cut to $[3.6]-[4.5]>-0.1$ recovers 98\% of quasars, although the
missing objects are predominantly high-$z$.  However, we must also
consider the level of contamination.  Figure~\ref{fig:s34compeff}
shows the tradeoff between (type 1) completeness and contamination.
Here we have made the simplifying assumption that any object lying
outside {\em either} the Lacy and Stern wedges are not quasars and
that objects lying {\em inside both} the Lacy and Stern wedges are
quasars.  For $[3.6]-[4.5]>-0.1$ the contamination fraction is nearly
60\%, though we caution that this number depends significantly on the
flux limit.  More importantly, if we desire to recover all of the
high-$z$ quasars, we must instead use $[3.6]-[4.5]>-0.4$, which has
over 85\% contamination.  On the other hand, our 6-D Bayesian
selection (which also uses only Channels 1 and 2) is 97.1\% complete
to known type 1 quasars (including high-$z$), with only 10\%
contamination, using the same criteria.




\section{Catalog}
\label{sec:cat}

Our catalog is presented in Tables~\ref{tab:cat} and \ref{tab:tab3},
where Table~\ref{tab:cat} describes the columns in
Table~\ref{tab:tab3}.  For the sake of completeness, we also tabulate
the 593 objects that were not selected by our Bayesian algorithms but
that otherwise meet our flux/magnitude criteria and are in {\em both}
the Lacy and Stern wedges; these objects are given in
Table~\ref{tab:tab4}.  The numbering of objects is common to
Tables~\ref{tab:tab3} and \ref{tab:tab4} and are sorted by right
ascension.

The first 25 columns in the data tables merely repeat the publicly
available optical and MIR information on these sources, see
\citet{aaa+08} and the references above for more information.
Columns 26--30 deal with object selection as discussed in
\S~\ref{sec:bayes}.  Columns 31--38 give photometric redshift
information as discussed in the next section.  Columns 39-41 provide
information on previous identification of these objects, whether
photometric \citep{rmg+09}, or spectroscopic (DR5x, \citealt{shr+07};
DR6x, \citealt{aaa+08}; T07x, \citealt{tim+07}; P06x;
\citealt{pap+05}).  For the spectroscopic identifications we have simply
repeated the classifications from the indicated references.

In all there are 5546 objects cataloged in Table~\ref{tab:cat}.  Note
that this number is similar to the number of objects in our quasar
training set.  This similarity is a result of our using the same flux
limit for both our training and test sets, while we could have
performed 6-D selection to much fainter limits (nearly 6$\times$ as
many objects).  As we consider our work here to be a
``proof-of-concept'', we save fainter 6-D quasar selection as an
exercise for the future after the current catalog has been more fully
validated with spectroscopic observations.  

Thus our procedure has essentially thrown out some wedge-selected
quasar candidates and has included some objects outside of the quasar
color wedges in the MIR.  That this is a worthwhile process can be
seen by noting that quasars make up 92\% of the known objects in among
our Bayesian selected targets (Table~\ref{tab:tab3}), while the
fraction of quasars among our rejected targets (Table~\ref{tab:tab4})
is only 29\%.  In addition, our Bayesian algorithm was shown to be
more robust than MIR-only color section over $3.5<z<5.0$
(Fig.~\ref{fig:zhisttrainall}), and is less contaminated at fainter
flux limits (Table~\ref{tab:tab2}).  These properties will be
particularly beneficial to the deeper census that can be performed
with objects detected in only the 2 bluest IRAC bandpasses.


\subsection{Photometric Redshifts}

\begin{figure}
\epsscale{0.9}
\plotone{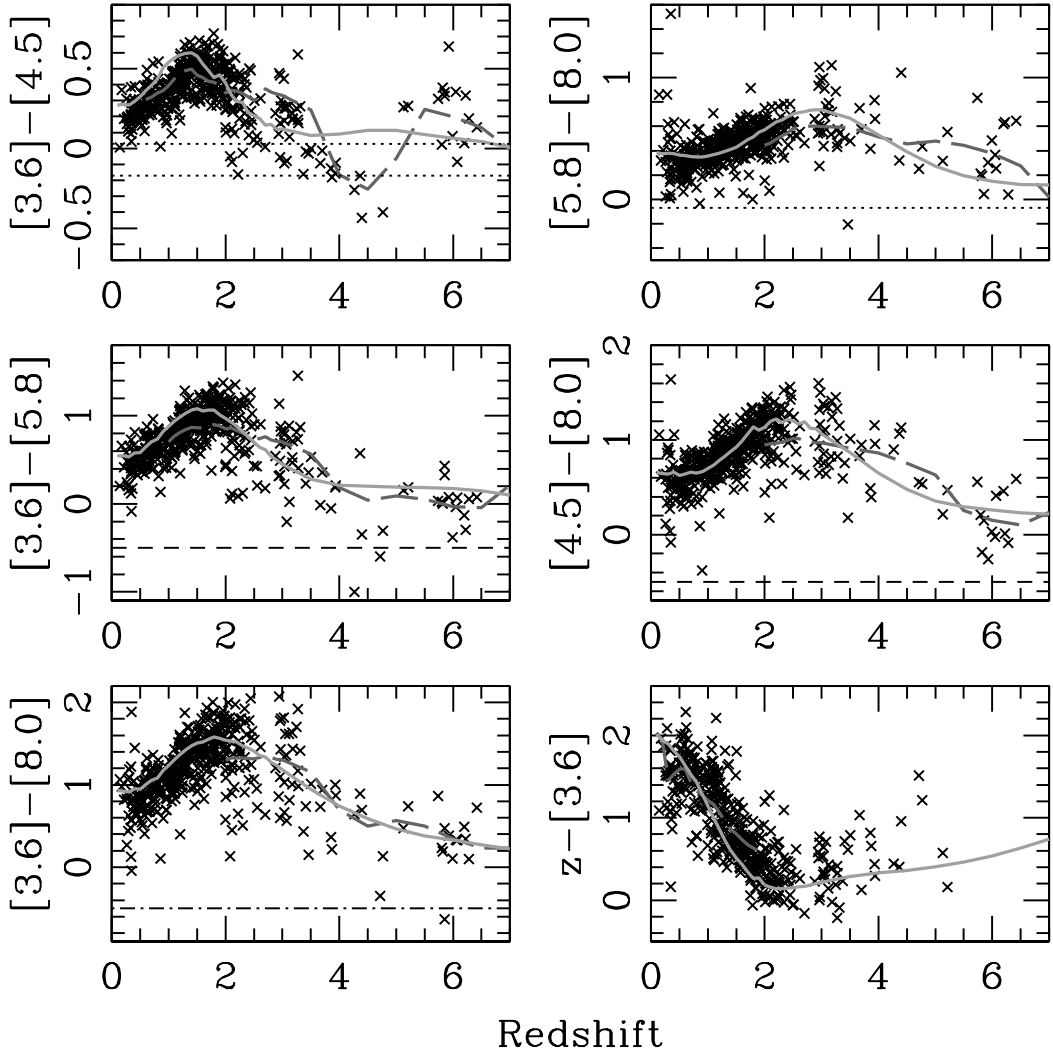}
\caption{MIR colors of known broad-line, type 1 quasars (black x's).
  The solid light gray and dashed dark gray curves show theoretical
  color redshift relation for a broad-band SED \citep{rls+06} and for
  the higher-resolution \citep{ghw06} IR spectral template,
  respectively, the latter more accurately representing the effects of
  the H$\alpha$ emission line.  In the upper two panels the dotted
  lines show the blue limits of the selection criteria for the Stern
  wedge (see Fig.~\ref{fig:iracstargal}, objects redder than these
  lines being inside the Stern wedge; the $[3.6]-[4.5]$ blue limit is
  a function of $[5.8]-[8.0]$, here only the extrema are plotted).  In
  the middle two panels the dashed lines shows the same for the color
  combination used in the Lacy wedge (objects redder than these lines
  being inside the Lacy wedge).  The bottom left-hand panel shows
  that, when coupled with morphology (see Fig~\ref{fig:iracstargal}),
  relatively complete and efficient selection is possible using only
  the $[3.6]-[8.0]$ color (dash-dot line).  The bottom right-hand
  panels shows the 1$\mu$m inflection induced color change in
  $z-[3.6]$, which is extremely beneficial for photo-z's.
\label{fig:czplotnew}}
\end{figure}

\begin{figure}
\plotone{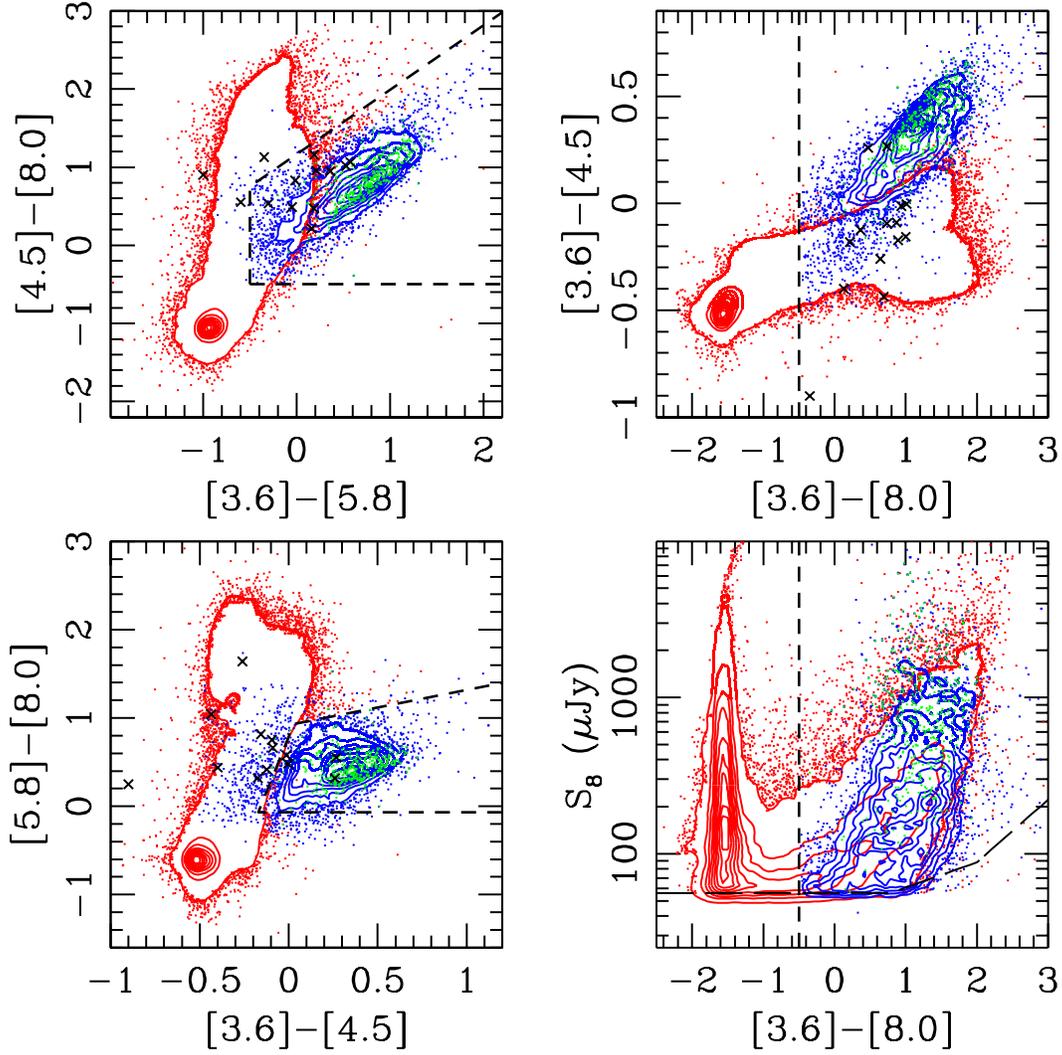} 
\caption{MIR color-color and color-mag distributions for the quasar (blue
  contours/dots) and non-quasar (red contours/dots) training sets.  Green
  points indicate known quasars; black crosses mark $z>3.5$ quasars.
  The long dashed line in the bottom right-hand panel shows our adopted
  $8.0\mu$m flux limit and how it is affected by the $3.6\mu$m and
  $5.8\mu$m limits.
\label{fig:iraccolorstrainlib}}
\end{figure}

\begin{figure}
\plotone{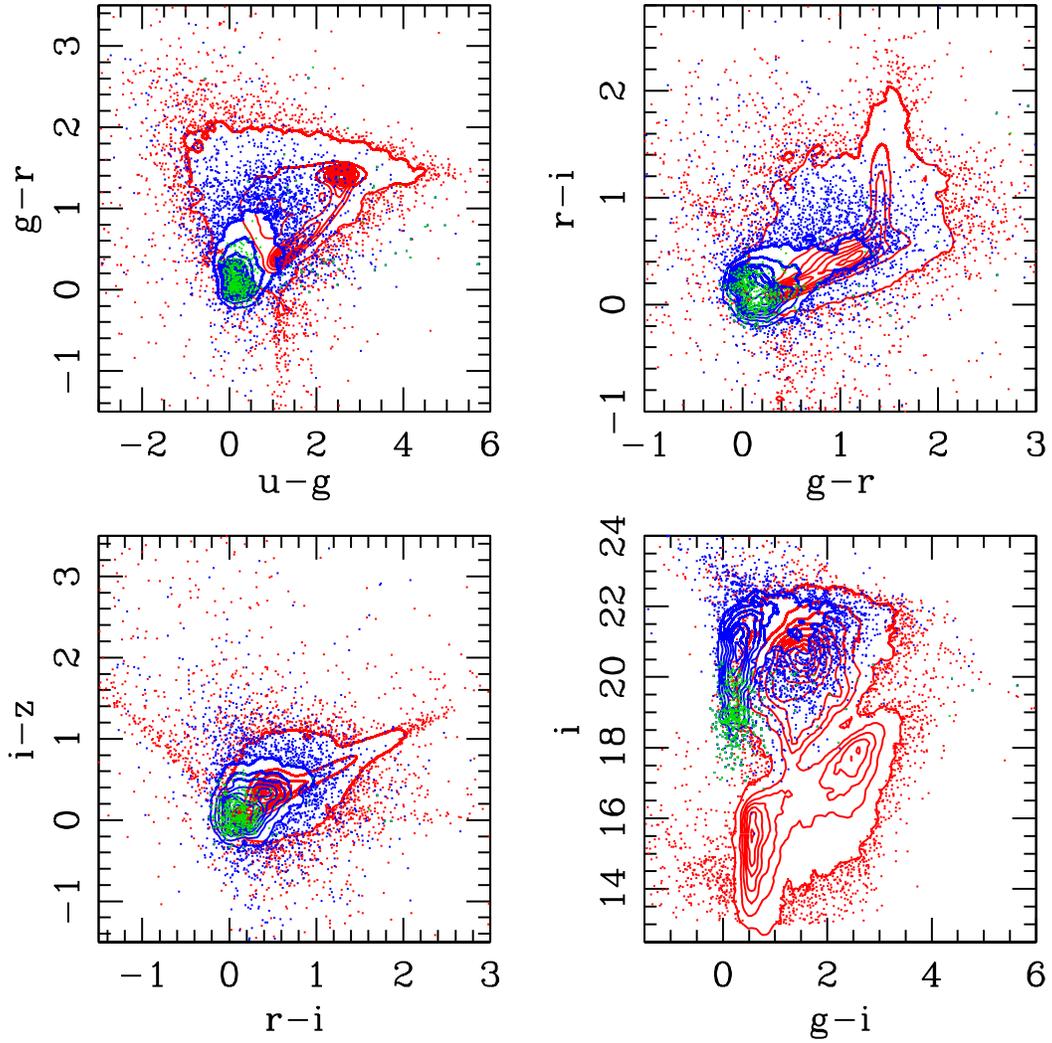} 
\caption{SDSS color-color and color-mag distribution for the quasar (blue
  contours/dots) and non-quasar (red contours/dots) training sets.  Green
  points indicate known quasars.
\label{fig:sdsscolorstrainlib}}
\end{figure}

\begin{figure}
\epsscale{0.9}
\plotone{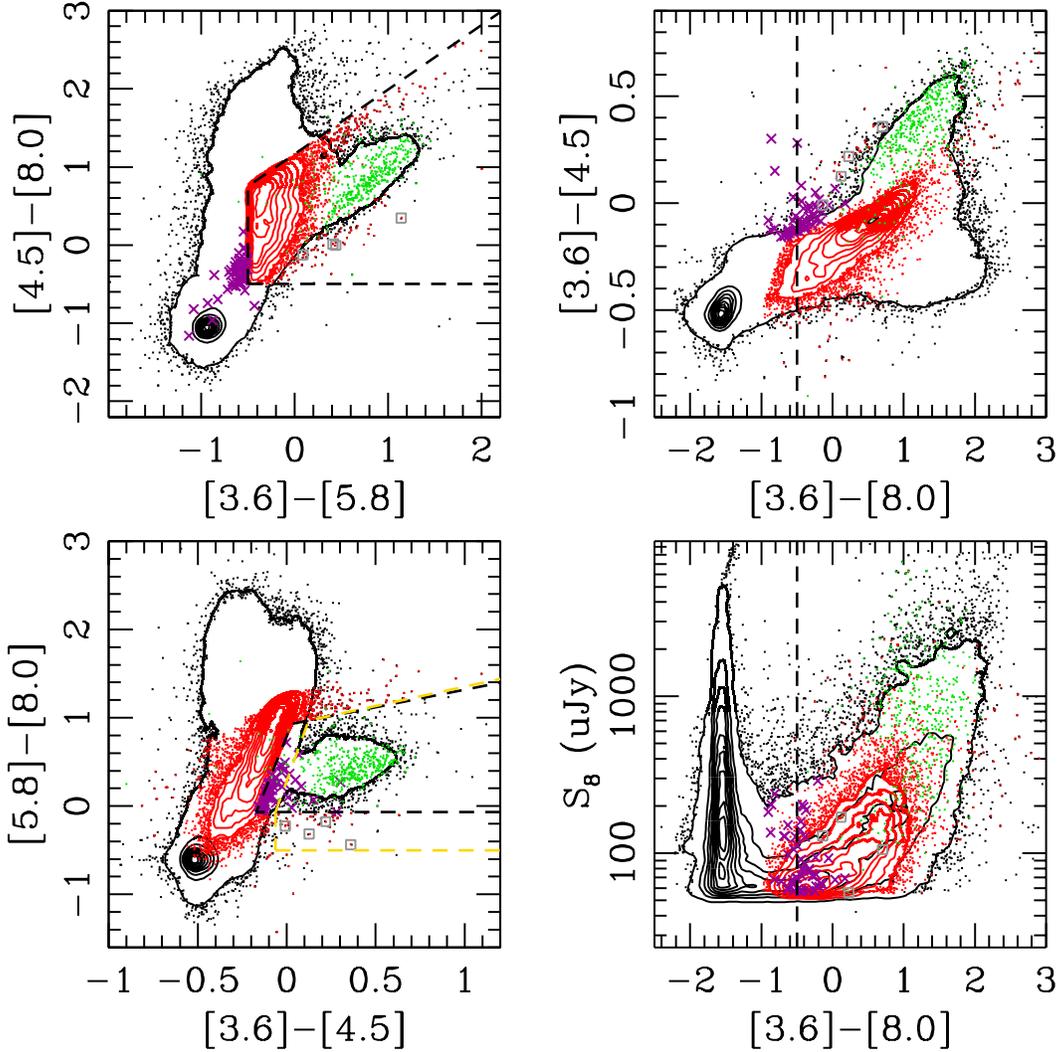} 
\caption{Comparison of the two standard MIR ``wedge'' color selection
  algorithms.  The top left panel shows that of \citet{lss+04} (dashed
  lines).  The bottom left panel that of \citet{seg+04}, again as
  dashed lines.  Black contours and dots are all sources detected in
  all 4 IRAC bands that have matches to SDSS sources.  Green points
  indicate known quasars.  Red contours/dots show objects inside the
  Lacy wedge, but outside the Stern wedge.  Purple crosses indicate
  objects in the Stern wedge, but outside the Lacy wedge.  The right
  panels show additional regions of color-color and flux-color space.
  The dashed line in the right-hand panels indicates a logical
  dividing line between stars and AGNs among point sources.  The
  region enclosed by yellow dashed lines in the lower left-hand panel
  indicates our modified Stern wedge, which is more conservative on
  the galaxy/AGN boundary in $[3.6]-[4.5]$, but is more inclusive to
  potential AGNs that have blue $[5.8]-[8.0]$ colors.  Open grey
  squares indicate the colors of 3 confirmed and one likely AGN
  resulting from spectroscopic follow-up of 9 random sources with
  $[5.8]-[8.0]<-0.07$.
\label{fig:wedgecomp}}
\end{figure}

\begin{figure}
\plotone{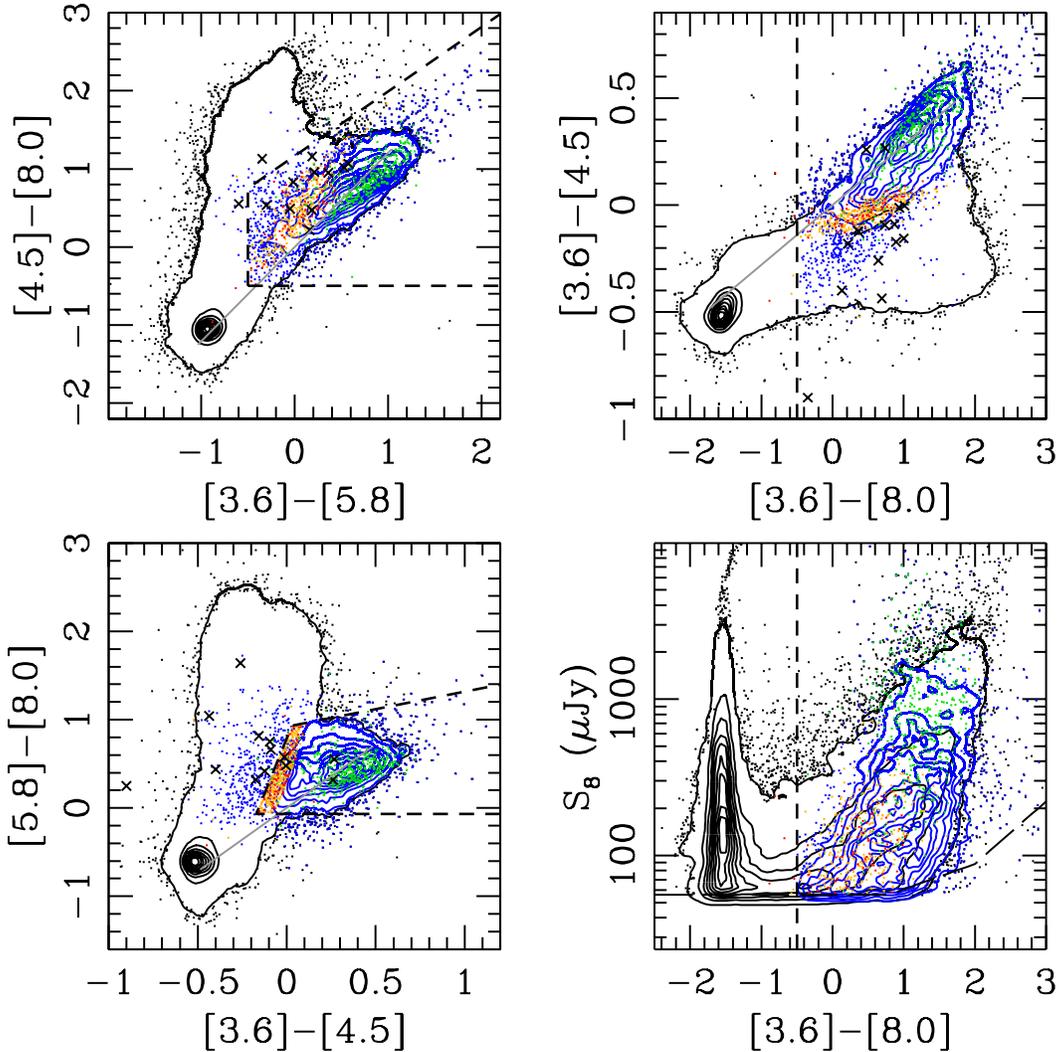} 
\caption{MIR colors/flux densities of Bayesian selected quasars.
  Black contours/dots show the full sample. Blue contours/dots show
  objects selected by both the 8-D and 6-D Bayesian algorithms.  Gold
  (red) points are objects selected only by the 8-D (6-D) algorithm.
  Known quasars are shown in green with $z>3.5$ quasars shown by black
  crosses.  Theoretical power-law colors ($-2<\alpha_{\nu}<2$) are
  given by the grey line (where we assume a delta function filter
  curve at the nominal effective wavelength).  Deviations from
  power-laws are seen in the Stern wedge due to the use of adjacent
  bandpasses where the small-wavelength structure can dominate the
  overall shape of the MIR SED.
\label{fig:iraccolorstest79Dclass}}
\end{figure}

\begin{figure}
\plotone{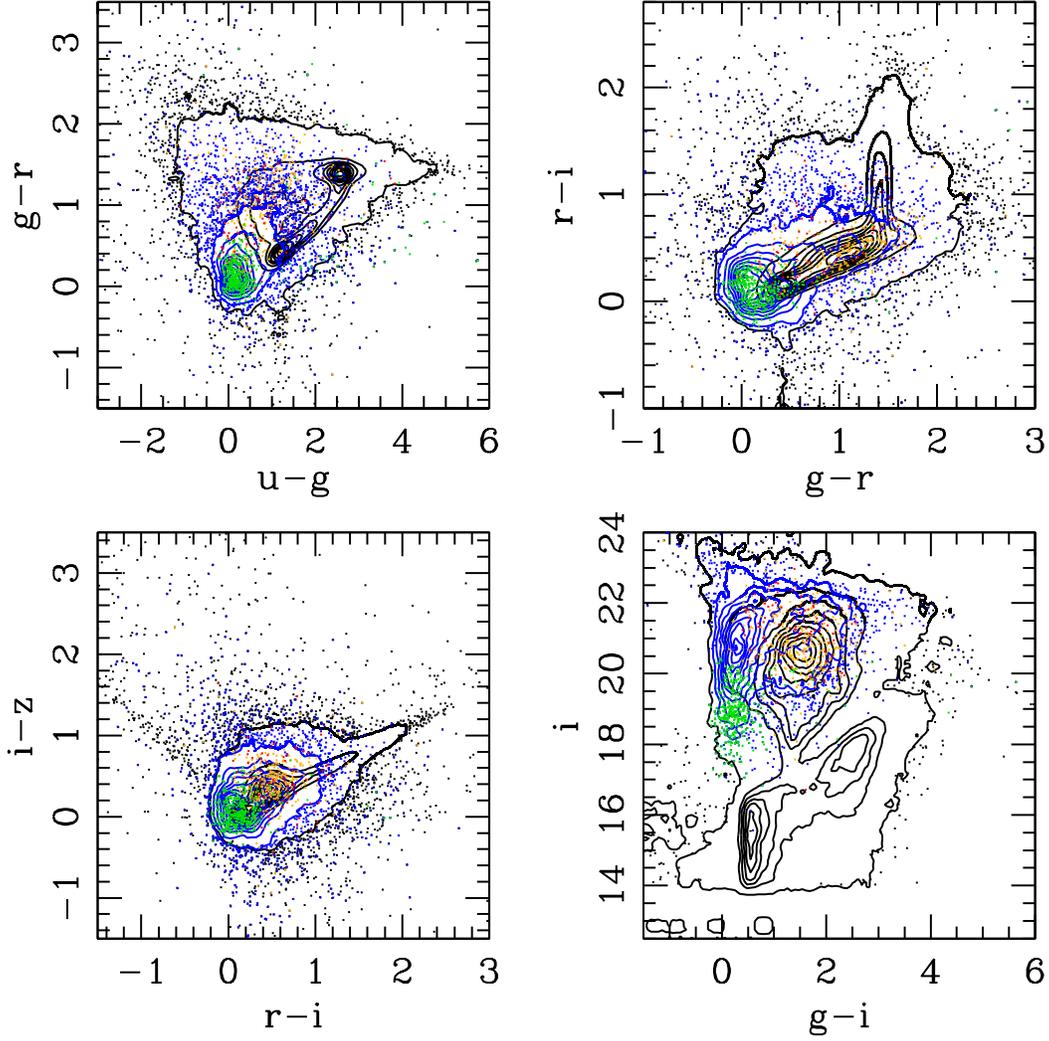} 
\caption{Optical colors/mags of Bayesian selected quasars.  Black
  contours/dots show the full sample. Blue contours/dots show objects
  selected by both the 8-D and 6-D Bayesian algorithms.  Gold (red)
  points are objects selected only by the 8-D (6-D) algorithm.  Again,
  known quasars are shown in green.
\label{fig:sdsscolorstest79Dclass}}
\end{figure}

\begin{figure}
\plotone{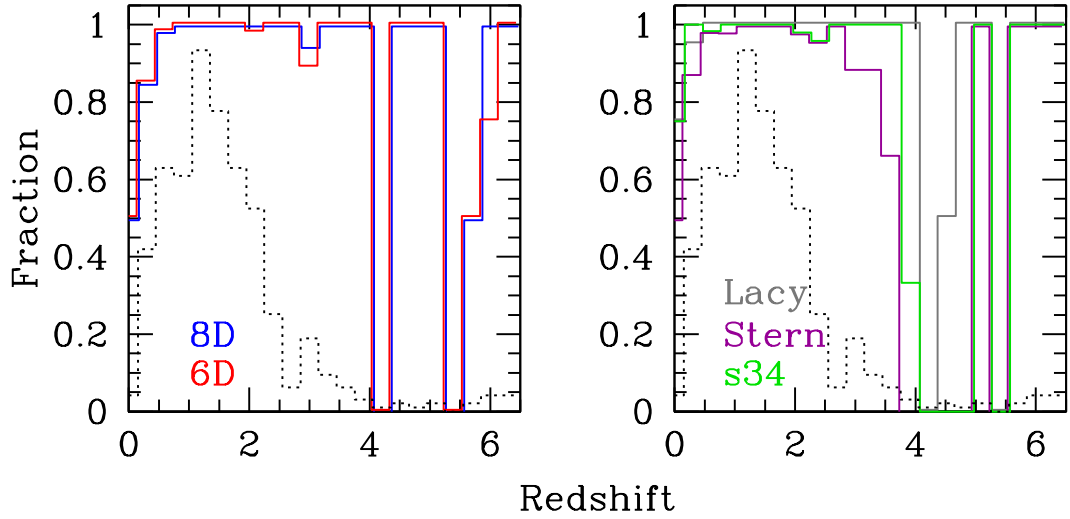} 
\caption{Fraction of known (type 1) quasars recovered.  {\em Left:} Our 8-D
  (blue) and 6-D(red) Bayesian selection algorithms.  {\em Right:} The
  Lacy wedge (gray), the Stern wedge (purple) and a $[3.6]-[4.5]$
  color cut (green).  The dashed black line in each panel indicates
  the number of known objects in each redshift bin. The peak at
  $z=1.2$ has 89 objects.  $z=5.4$ has 0 objects and $z=4.2$ and
  $z=4.8$ have only one each, thus statistics are poor in these bins.
\label{fig:zhisttrainall}}
\end{figure}

\begin{figure}
\plotone{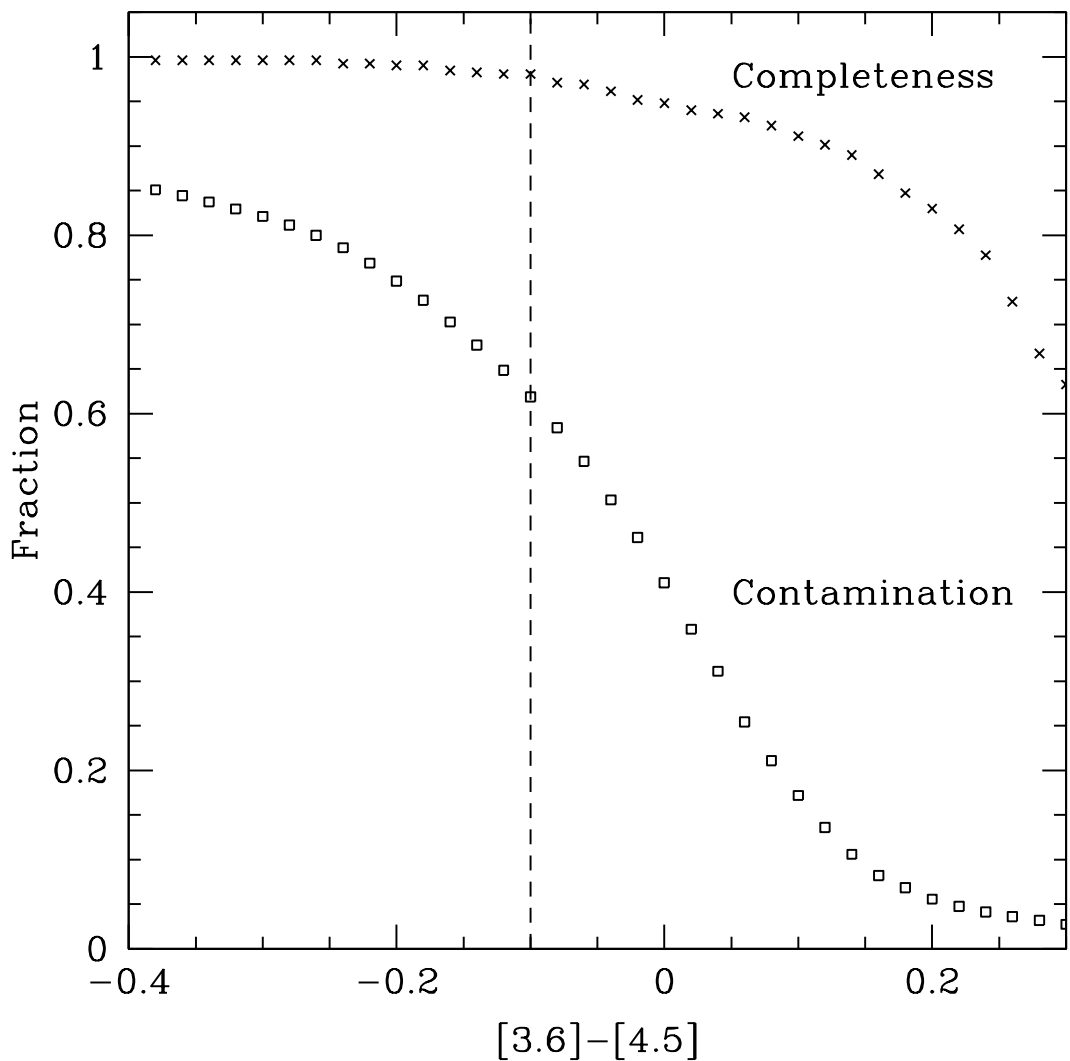} 
\caption{Completeness vs.\ contamination for quasar selection based on
  a cut in $[3.6]-[4.5]$ alone.  Crosses indicate the fraction of
  known (type 1) quasars recovered as a function of $[3.6]-[4.5]$
  color.  For $z\lesssim2$, type 2 quasars have similar $[3.6]-[4.5]$
  colors and should be equally included/excluded.  The squares show
  the contamination fraction for the same color cut.
\label{fig:s34compeff}}
\end{figure}

While the errors on the IRAC flux densities ($\sim10$\% for IRAC [but
  see \citealt{hcs+08}] vs.\ $\sim2$\% for SDSS) are generally too
large to permit accurate MIR-only photometric redshift estimation, the
combination of SDSS and IRAC photometry allows for considerable
leverage in estimating redshifts of AGNs.  We have updated the
algorithm described by \citet{rws+01} and \citet{wrs+04} to operate on
any number of color dimensions.  In essence, quasar photometric
redshift estimation relies on the small, but distinct, color changes
produced as broad emission lines move through photometric bandpasses.
At $z\gtrsim2.3$, Lyman-$\alpha$ forest absorption mimics the Balmer
break that is so useful in reducing galaxy photometric redshift
estimates to the 2\% level \citep[e.g.,][]{ccs+95}.  However, quasar
emission lines are generally strong enough that they also produce
measurable features in the color-redshift relations that can be used
to estimate photometric redshifts (photo-$z$'s).  The mean SDSS colors
as a function of redshift were shown most recently by \citet{shr+07},
while the redshift dependence of IRAC colors for SDSS quasars was
given by \citet{hpp+04} and \citet[][Fig.~3]{rls+06}; see also
Figure~\ref{fig:czplotnew}.

\begin{figure}
\plottwo{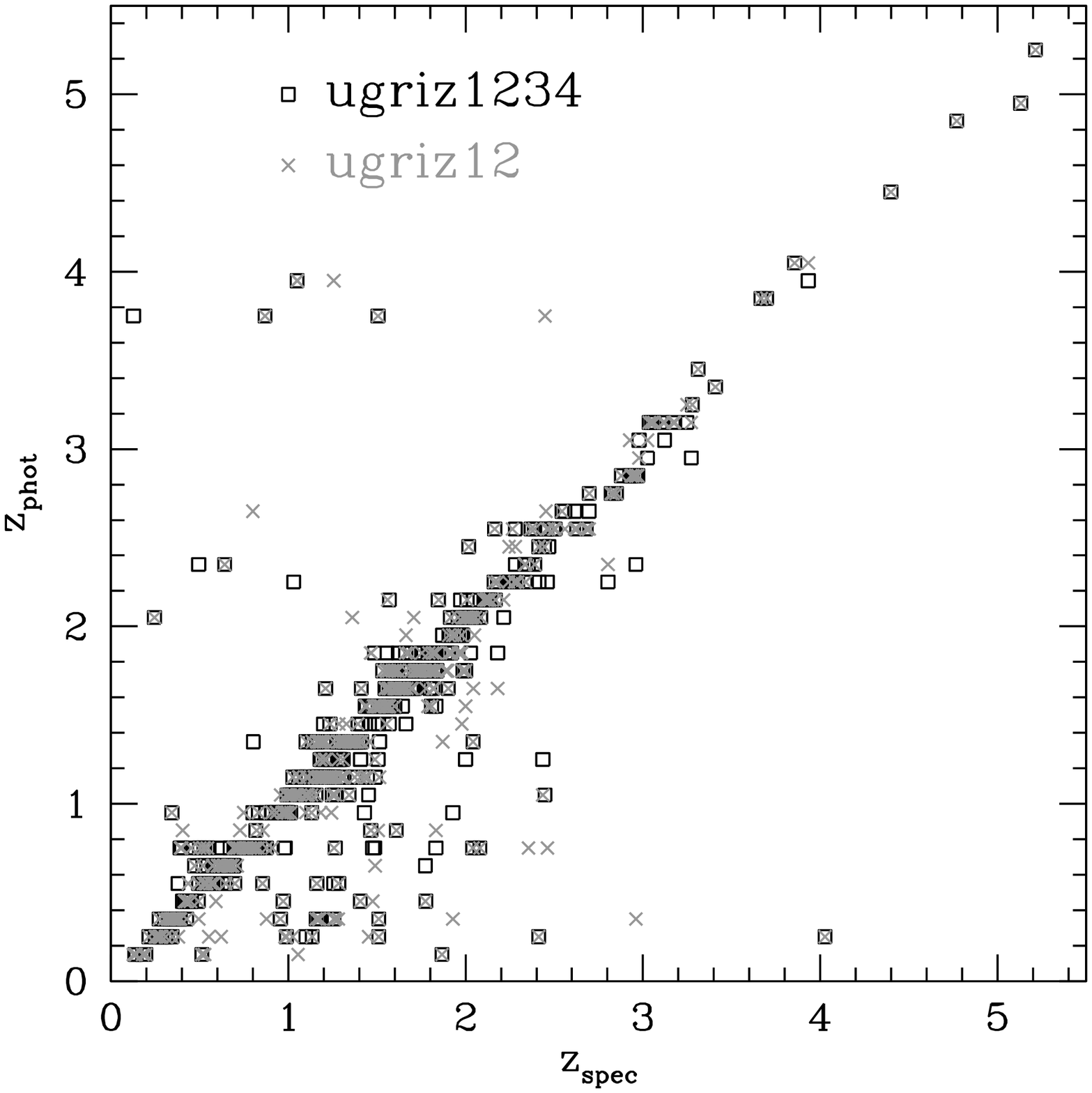}{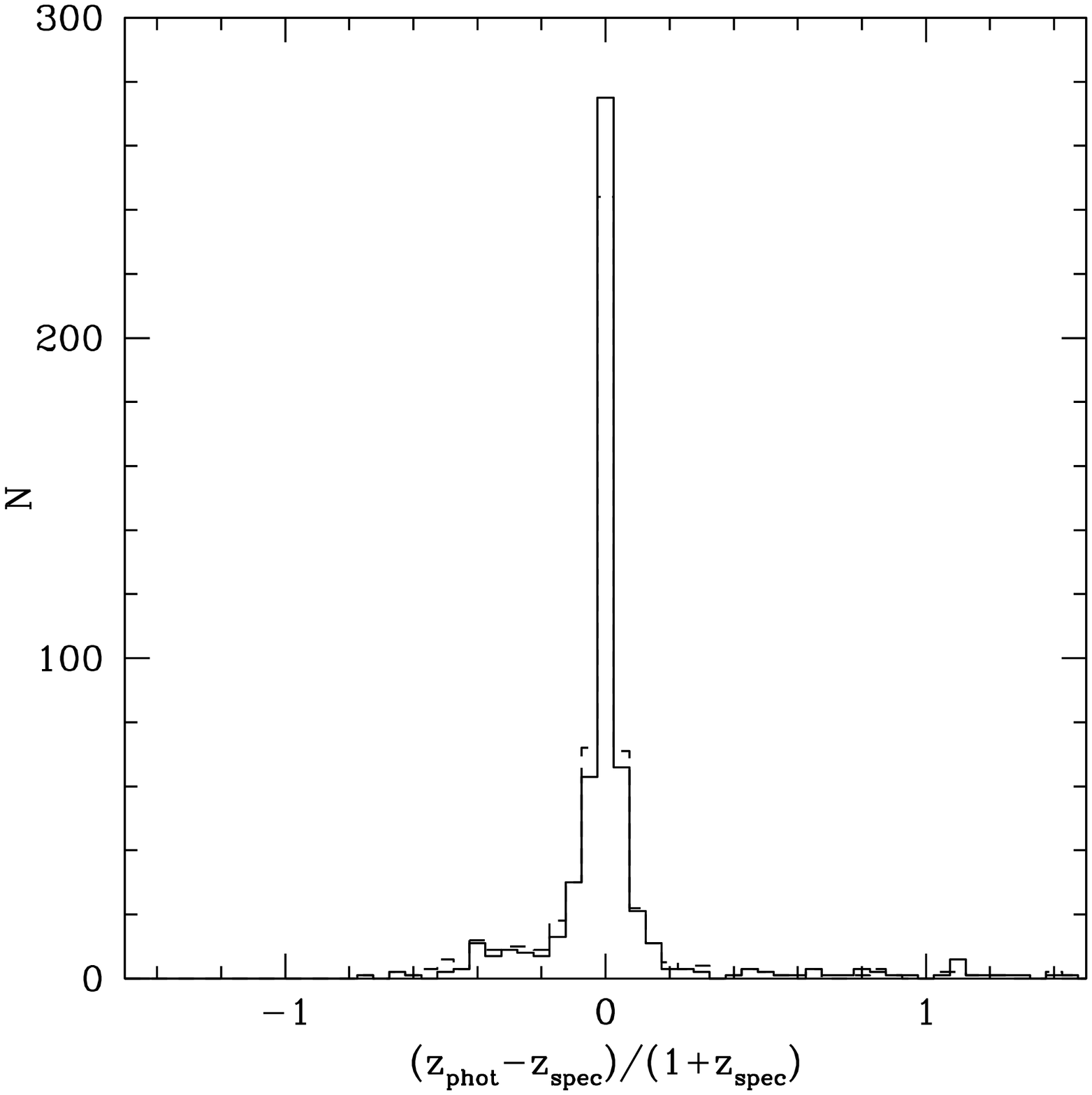}
\caption{{\em Left:} Photometric vs.\ spectroscopic redshift for 399 known
  quasars.  Open black squares show photometric redshifts determined
  from 8 colors.  Grey crosses show photometric
  redshifts determined from 6 colors.  {\em Right:} Histogram of the
  fractional error in photometric redshift.  8-color photo-z results
  are shown by the solid line; 6-color by the dashed line.  Little
  accuracy is lost by going from 8 to 6 colors.  The bin size is 0.05 in redshift.
\label{fig:zzplot}}
\end{figure}

Using the IRAC-detected, spectroscopically-confirmed quasars noted
above, we have updated the quasar color-redshift relations and
computed the photometric redshifts for all of our quasar candidates.
In addition, Table~\ref{tab:cat} provides not only the most likely
photometric redshifts, but also a range and the probability that the
actual redshift is within that range.  We compute photometric
redshifts both using both 8 colors and 6 colors (dropping the 2
reddest IRAC bandpasses to simulate the {\em Spitzer} warm mission
data).  In the left-hand panel of Figure~\ref{fig:zzplot}, we show
both the 8-D and 6-D photometric redshifts versus spectroscopic
redshifts.  The 6-color photo-z's are not significantly worse than the
8-color photo-z's.  This is perhaps not surprising given that most of
the information that comes from adding the IRAC bandpasses is provided
by the $z-[3.6]$ color which, due to the 1 micron inflection, spans an
impressive 2 magnitudes over $0<z<2$ (see the bottom right-hand panel
in Fig.~\ref{fig:czplotnew}) and serves to break most of the redshift
degeneracies seen when computing photometric redshifts from SDSS
photometry alone.  Within $\pm 0.3$ in redshift, we find that both the
8-color and 6-color photo-z's are accurate 93.6\% of the time.  In
addition, the majority of the photo-z's are considerably more accurate
than $\pm0.3$; 82--84\% are accurate to $\pm 0.1$ in redshift.  This
accuracy is illustrated in the right-hand panel of
Figure~\ref{fig:zzplot}, where we show a histogram of the fractional
redshift errors.  Most of the outliers are fainter objects with
$i>19.1$, but even for those the fraction of 6-color photoz's accurate
to $\pm0.3$ is 92\%.

In principle, any deep optical imaging data can be used to determine
photometric redshifts, but in practice the SDSS filter set is nearly
ideal for photometric redshifts of quasars due to the more ``top-hat''
nature of the SDSS filters than the traditional
Johnson-Morgan/Kron-Cousins filters.  We emphasize that
our goal here is primarily to find type 1 quasars for which our
photometric redshift algorithm should work quite well.  However, at
faint flux limits and/or larger host galaxy contribution, our
templates fail to yield accurate photo-z's --- necessitating more
careful photometric redshift techniques.  For example, if the
multi-wavelength coverage is large enough (e.g., UV, optical, near-IR,
and MIR), it is also possible to determine photometric redshifts
through SED template fitting \citep[e.g.,][]{bba+06,rbo+08,shi+08} which
also enables simultaneous photometric redshift estimation for inactive
galaxies and type 2 quasars.

\subsection{Bulk Properties}

\begin{figure}
\plotone{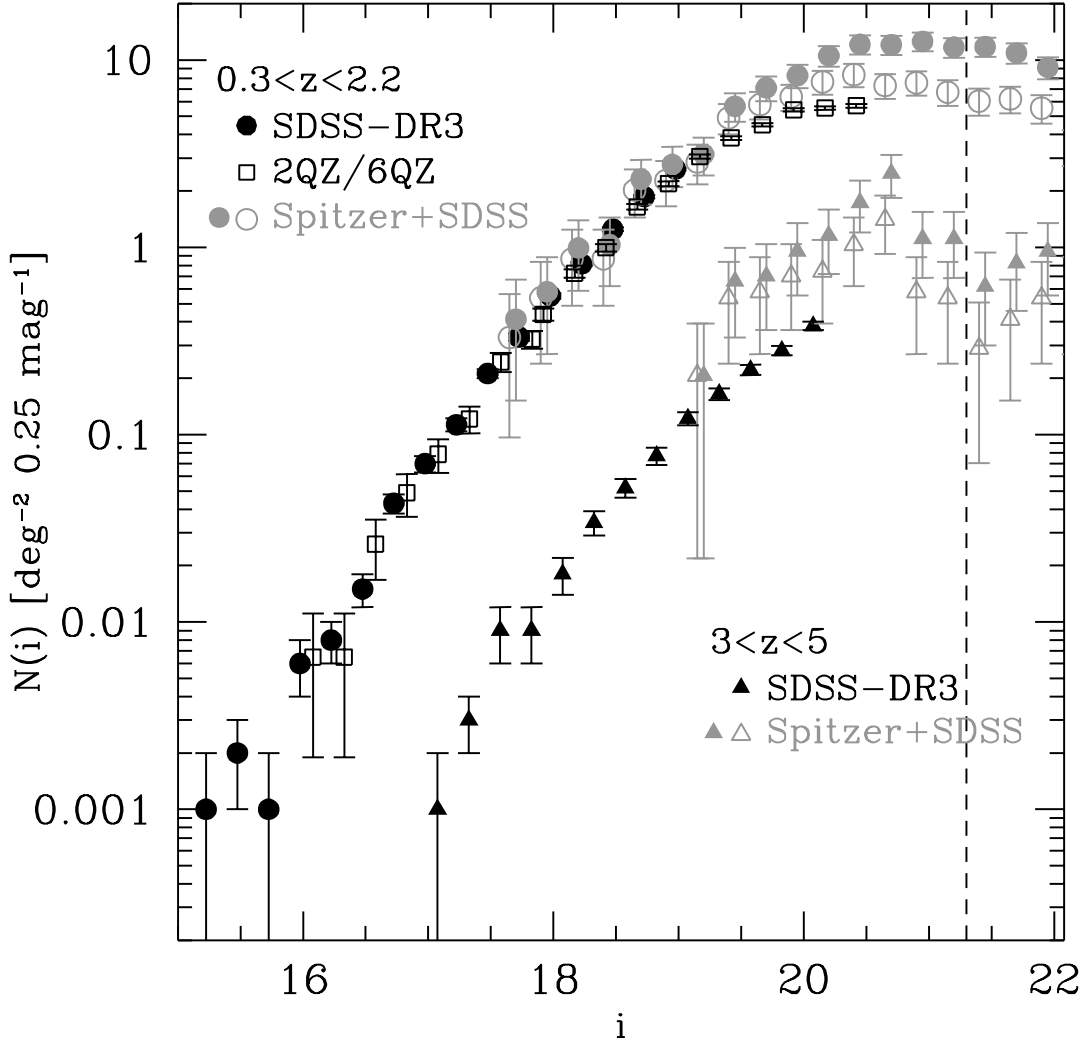} 
\caption{Number counts ($i$-band) of {\em Spitzer}+SDSS selected quasars as
  compared to SDSS and 2QZ number counts.  Two redshift bins are
  shown: $0.3<z<2.2$ as circles/squares and $3<z<5$ as triangles.  Our
  new results are shown for two extreme cases: objects selected by
  either the 6-D or 8-D algorithms and with photometric redshift
  probability larger than 0.5 (closed points) and for objects selected
  by both the 6-D and 8-D algorithms and with photometric redshift
  probability larger than 0.8 (open points).  For objects brighter
  than the nominal SDSS flux limit of $i=21.3$, these extremes should
  bracket the true values.  Points are truncated at the bright end
  where the errors become large due to the lack of area.
\label{fig:nmi}}
\end{figure}

With the aid of photometric redshifts, we can examine other properties
of the catalog.  Figure~\ref{fig:nmi} shows the number counts of our
sample as compared to the SDSS-DR3 \citep{rsf+06} and 2QZ results
\citep{csb+04}.  Two extreme cuts on our catalog are shown to give the
reader the range of possible values as objects get fainter and
classification (and redshift estimation) become less robust.  The
number counts are just slightly above those of the SDSS spectroscopic
sample for relatively bright $z<2.2$ sources, which is consistent with
the fact that fully or partially obscured quasars in our sample have
not been corrected for internal extinction (and thus will be shifted
to fainter bins).  As such, even in the presence of a significant
population of obscured sources, we would not expect the number counts
as a function of observed magnitude to be much higher than for the
SDSS spectroscopic sample in this range.  For $3<z<5$ quasars, we see
a marked increase in our sample as compared to the SDSS spectroscopic
sample.  This is likely a combination of the inclusion of obscured
quasars that are intrinsically brighter and also due to dust reddened
quasars having a greater tendency to (incorrectly) have higher
photometric redshifts (since high-$z$ quasars have redder colors at
short wavelengths).

\begin{figure} 
\epsscale{0.9}
\plotone{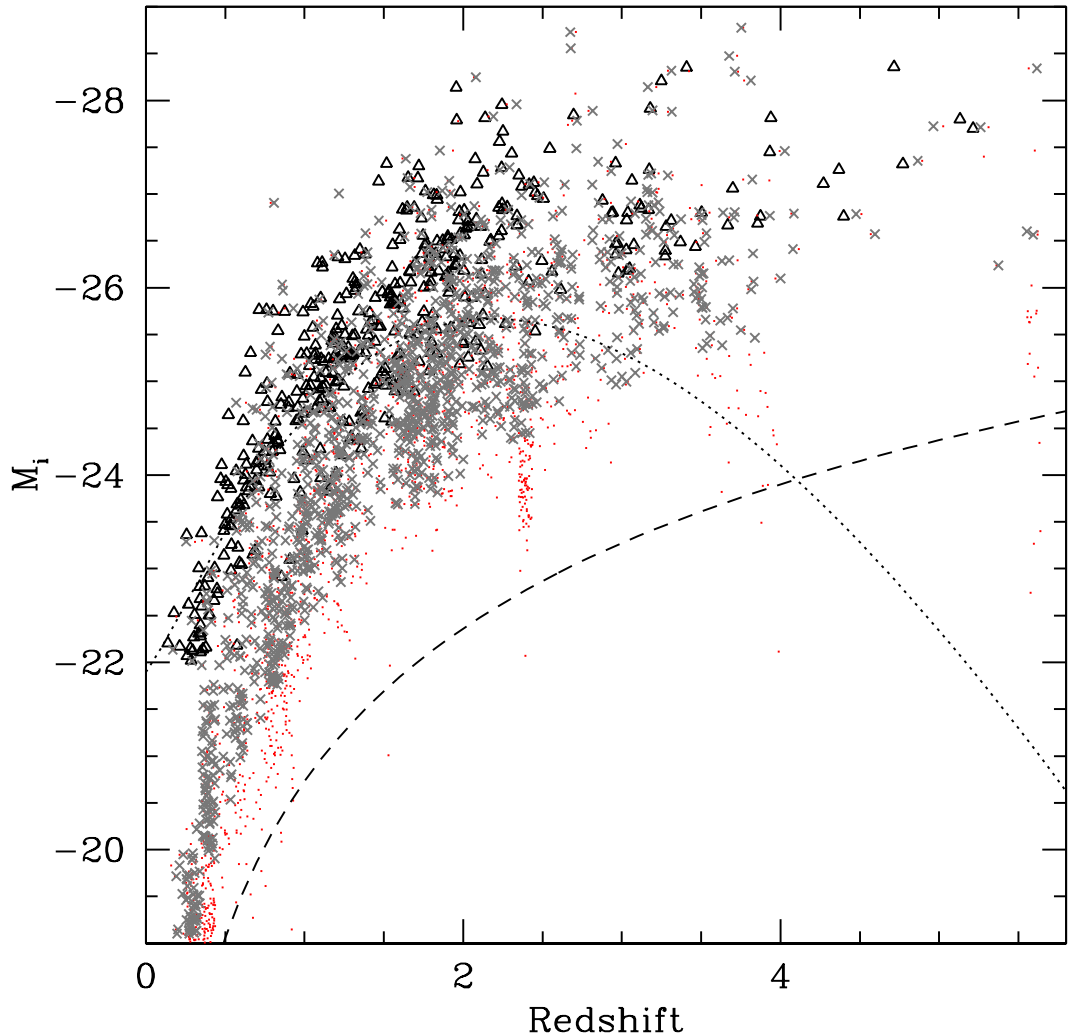} 
\caption{Absolute $i$-band magnitude vs.\ redshift.  Known SDSS
  quasars are shown as black triangles.  Robust MIR+optical quasar
  candidates are shown by grey crosses.  These are objects detected by
  both the 8-D and 6-D Bayesian methods, that have point-like SDSS
  morphologies, $i<21.3$ (the SDSS flux limit), and photometric
  redshift probabilities greater than 90\%.  The remaining quasar
  candidates are plotted as red dots and more likely to include
  contaminants and erroneous photometric redshifts.  The dotted black
  line shows the approximate division between the bright and faint
  ends of the quasar luminosity function (i.e., $L^*_Q$) as derived
  from \citet{hrh07}.  This demonstrates that SDSS quasars only probe
  the bright end of the QLF, while adding MIR information enables us
  to probe the faint end of the QLF.  The dashed black line indicates
  the depth that can be reached if optical+MIR selection could be
  performed to $i=23$ (e.g., SDSS Stripe 82).  At that depth, the
  faint end of the QLF can be probed to nearly $z=4$, and there is
  sufficient dynamic range in luminosity at $z<3$ to determine the
  luminosity dependence of quasar clustering as a function of
  redshift.
\label{fig:zmplot}} \end{figure}

Figure~\ref{fig:zmplot} shows the distribution of our objects in the
absolute magnitude vs.\ redshift plane, and demonstrates how this
sample can be used to help break luminosity-redshift degeneracies
inherent to any flux-limited survey.  For example, $z\sim3.5$ quasars
can now be compared to $z\sim2.5$ quasars at the same luminosity.
Note that our $f_{8.0\mu{\rm m}}<56\,\mu$Jy restriction corresponds
roughly to 15--20$\,\mu$Jy at 3.6$\mu$m, which is about a factor of 3
shallower than (5-$\sigma$) SWIRE-depth (120s) IRAC data, thus there
is room for further improvements in dynamic range.  Similarly, the
dashed line shows the improvement that can be expected from using the
SDSS southern equatorial stripe (``Stripe 82''), which has up to 100
epochs of SDSS imaging data, yielding a co-added flux limit of
$i\sim23$ \citep{alm+06}.  For luminous sources ($M_i\lesssim-23$) the
central engine dominates over the host galaxy, but for less luminous
sources (with $z\lesssim2.5$) we will have to account for a more
significant host galaxy contribution.  The need for more dynamic range
in luminosity is illustrated by the dotted black line which traces the
"break" luminosity in the quasar luminosity function (QLF) using the
multi-wavelength bolometric determination of \citet{hrh07}.
As discussed in the introduction, the ability to determine the
luminosity dependence of quasars clustering (particularly at $z>2.5$),
and the slope evolution of the faint end of the quasar luminosity
function are among the most important near-future constraints on
feedback models of galaxy evolution.  Our selection algorithm will
help to enable the creation of a quasar sample that can be used the
address these issues.

\section{Obscured Quasars}
\label{sec:type2}


Note that, while morphology was used in the creation of the training
sets, no optical morphology information is used in the actual Bayesian
selection.  Thus we expect that our selection includes some type 2
quasars in addition to the type 1 quasars.  This simply reflects the
fact that 
type 2 quasars have similar colors in the MIR as type 1 quasars.  This
is true for redshifts low enough ($z\la 2$) that the IRAC bands probe
thermal emission of circumnuclear dust. At higher redshifts, as
rest-frame optical and NIR emission moves into the IRAC passbands, MIR
colors of type 2 quasars deviate significantly from those of type 1
quasars; our procedure is not expected to be sensitive to such
objects.

\begin{figure}
\plotone{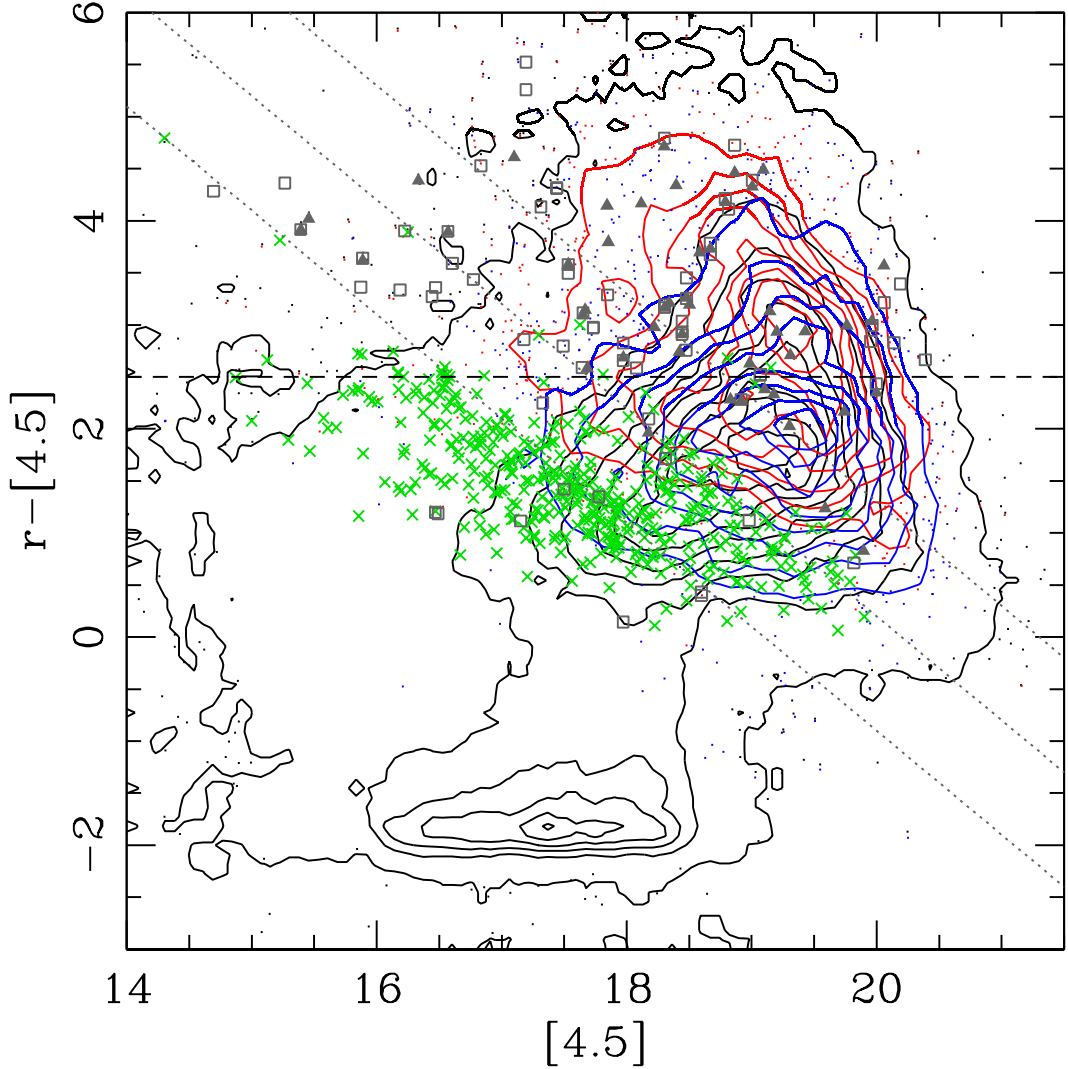} 
\caption{Optical/MIR color-mag relationship showing the separation of
  type 1 and type 2 quasars in this plane, similar to \citet{hjf+07}.
  Known type 1 quasars are shown by green crosses.  Known type 2
  quasars are open grey squares; filled gray triangles indicate those
  recovered by our algorithm.  Our point source quasar candidates are
  shown in blue.  Extended quasar candidates are shown in red.  The
  separation between these populations is similar to that seen by
  \citet{hjf+07} for type 1 and type 2 AGNs using a dividing line of
  $r-[4.5]=2.5$.  Note however, that optical magnitude and redshift
  must be considered to some extent.  The dotted lines show $i=19.1,
  20.2$, and $22$ from bottom to top and demonstrates that the
  apparent diagonal locus of type 1 sources is artificial.
\label{fig:hickoxall}}
\end{figure}

As \citet{hjf+07} have shown, type 1 and type 2 quasars are relatively
well separated in optical-to-MIR flux ratio; we examine this
distribution for our sources.  Figure~\ref{fig:hickoxall} shows the
$r-[4.5]$ vs.\ $[4.5]$ color-mag distribution of our Bayesian-selected
quasar candidates with known type 1 (green) and type 2 (gray) quasars.
While \citet{hjf+07} use luminosity in their plots, our photometric
redshifts will be in error for type 2 quasars, so we have made the
plot in flux units.  This choice has no effect on the vertical axis
which is used by \citet{hjf+07} to separate type 1 and type 2 quasars
(modulo the change in units used herein), but it does cause a
different distribution along the horizontal axis.  Point sources
(blue) and extended sources (red) have well-separated mean values that
are bifurcated along $r-[4.5]$, similar to the type 1/2 dividing line
advocated by \citet{hjf+07}.  This comparison suggests that our sample
contains a significant number of type 2 quasars --- despite the fact
that we have only attempted type 1 selection ({\em and have required
  matching to the relatively shallow single-epoch SDSS photometric
  catalog}).

While our sample is certainly not complete to type 2 quasars, we
note that type 2 quasars are more likely to have extended morphologies
and that Figure~\ref{fig:hickoxall} shows a ``valley'' between point
(peak $r-[4.5]\sim2$) and extended (peak $r-[4.5]\sim3$) morphology
quasar candidates at a color of $r-[4.5]\sim2.5$.  As such, the most
logical place to look for type 2 quasars among our quasars candidates
would be those extended sources with $r-[4.5]>2.5$ of which there are
1252 in all.  In all 28 of 47 (60\%) known type 2 quasars that we
recover are extended sources with $r-[4.5]>2.5$.  Comparing to the
type 2 quasars cataloged by \citet{tim+07} in the COSMOS field, our
algorithm recovers 20 of 66 (30\%) with $i<21.3$ and $f_{8.0\mu{\rm
    m}}<56\,\mu$Jy.

However, this dividing line in $r-[4.5]$ is not absolute and
the optical magnitude and redshift play a significant role in the
location of objects in this diagram.  Nevertheless, it seems likely
that $r-[4.5]$ can be used as a crude diagnostic of type 1 vs.\ type 2
AGNs as suggested by \citet{hjf+07}.  On the other hand, point sources
with $r-[4.5]\le2.5$ are quite likely to be type 1 quasars and there
are 2536 such objects in our catalog.

In addition to quasars whose central engines are fully obscured in the
optical, there also exist quasars that are simply heavily reddened,
but still exhibit broad-line emission features that are characteristic
of type 1 quasars.  Predictions of the size of this population range
from $\sim15$\% \citep{rhv+03} to $\sim60$\% or more
\citep[e.g.,][]{whb+03,ghw+07}.  Most recently \citet{mhw+08} have
used UKIDSS\footnote{http://www.ukidss.org/} data to argue that the
fraction of type 1 quasars missing from $i$-band selected surveys
(i.e., SDSS) is $\sim$30\%.  Recent work suggests that some of this
dust reddening may come from the host galaxy
\citep[e.g.,][]{mgt98,smh+07,dck+07,pwh+08} rather than the putative
dusty torus.  Thus a complete census is needed to fully understand the
demographics of black hole fueling (and how it affects galaxy
evolution), and these quasars represent a important, but under-studied
population.  As with type 2 quasars, our emphasis on unobscured type 1
quasars should allow for more complete selection of quasars that have
been extincted or reddened out of purely optically selected samples.
Indeed, we are able to recover 6 of the 7 reddened type 1 quasars in
the XFLS area cataloged by \citet{lps+06}.

\section{Conclusions}
\label{sec:conclusions}

In this paper we present a method to select type 1 quasars from a
combination of optical and MIR photometric data.  The method is based
on Bayesian analysis techniques in multi-dimensional MIR+optical color
space.  We demonstrate that our algorithm presents a significant
improvement over MIR-only and optical-only selection procedures.  Both
the completeness of selection (i.e., the percentage of true quasars
recovered by the method) and the robustness of selection (i.e., the
percentage of objects recovered that are quasars rather than
contaminants) are increased.  In all, we catalog 5546 quasar
candidates detected in all four bands of IRAC in $\sim 24$ deg$^2$
($>200$ deg$^{-2}$), yielding a factor of $\sim20$ increase in density
compared to that of SDSS spectroscopic quasar catalog.  Relaxing our
requirement for detections in IRAC Channels 3 and 4 would increase the
quasar density by more than a factor of 5 ($\sim1000$\,deg$^{-2}$).

Comparison with existing samples shows that the catalog is more than
95\% complete to known type 1 quasars at all redshifts.  By combining
the 5 SDSS and the 4 IRAC bandpasses, we recover the $2.5<z<3.0$
quasars lost by optical selection due to contamination by stars and
the $3.5<z<5.0$ quasars lost by the MIR selection due to contamination
by star-forming galaxies.  Furthermore, we find that combining optical
and MIR data allows selection of quasar candidates to much fainter
fluxes than those afforded by the MIR cuts currently in use in the
literature.  At the same time, working in the optical+MIR color-color
space greatly helps with rejecting contaminants (stars and inactive
galaxies) in an efficient manner.

Inclusion of MIR data significantly improves photometric redshift
estimation for type 1 quasars.  The fraction of quasars with redshifts
within 0.3 of the true values increases from $\sim 80\%$ for the
optical-only photo-$z$ to $\sim94$\% for optical+MIR photo-$z$.  Much of
this improvement is due to a rapid change of $z-[3.6]$ color of
quasars as a function of redshift at $z\la 2$.

We demonstrate that removing the two longest wavelength IRAC channels
has little detrimental effect on the selection procedure and on the
quality of photometric redshift estimates.  Therefore our method can
be successfully used on data collected during the warm extension of
the {\it Spitzer} mission.  
A wide-area sample of overlapping deep optical and MIR data would make
groundbreaking contributions to our understanding of quasar feedback
and the evolution of galaxies by breaking the redshift-luminosity
degeneracy inherent to current quasars surveys (Fig~\ref{fig:zmplot}).

Although the algorithm was primarily designed to be complete for  
selection of type 1 quasars, it is also sensitive to at least some  
type 2 quasars. This is possible because MIR colors of low-redshift  
quasars are dominated by the thermal emission of circumnuclear dust  
and are therefore similar for type 1 and type 2 objects. We estimate  
that as many as 1200 of our quasar candidates are type 2 quasars.  
Although our procedure is not complete for type 2 quasars, our work  
lays the foundation for identification of type 2 quasars using modern  
statistical methods.

\acknowledgments

GTR acknowledges support from an Alfred P. Sloan Research Fellowship,
a Gordon and Betty Moore Fellowship in Data Intensive Sciences, and
NASA grants NNX06AE52G and 07-ADP07-0035.  We thank Michael Strauss
for critical review of the paper, and Ryan Riegel for help with the
non-parametric classificaiton algorithm.  Funding for the SDSS and
SDSS-II has been provided by the Alfred P. Sloan Foundation, the
Participating Institutions, the National Science Foundation, the
U.S. Department of Energy, the National Aeronautics and Space
Administration, the Japanese Monbukagakusho, the Max Planck Society,
and the Higher Education Funding Council for England. The SDSS is
managed by the Astrophysical Research Consortium for the Participating
Institutions. The Participating Institutions are the American Museum
of Natural History, Astrophysical Institute Potsdam, University of
Basel, Cambridge University, Case Western Reserve University,
University of Chicago, Drexel University, Fermilab, the Institute for
Advanced Study, the Japan Participation Group, Johns Hopkins
University, the Joint Institute for Nuclear Astrophysics, the Kavli
Institute for Particle Astrophysics and Cosmology, the Korean
Scientist Group, the Chinese Academy of Sciences (LAMOST), Los Alamos
National Laboratory, the Max-Planck-Institute for Astronomy (MPIA),
the Max-Planck-Institute for Astrophysics (MPA), New Mexico State
University, Ohio State University, University of Pittsburgh,
University of Portsmouth, Princeton University, the United States
Naval Observatory, and the University of Washington.

This work is based [in part] on archival data obtained with the
Spitzer Space Telescope, which is operated by the Jet Propulsion
Laboratory, California Institute of Technology under a contract with
NASA. Support for this work was provided by an award issued by
JPL/Caltech (\#1290740).  Part of this work is based on observations
obtained with {\em XMM-Newton}, an ESA science mission with
instruments and contributions directly funded by ESA Member States and
the USA (NASA).  Some data presented here were obtained at KPNO, a
division of the NOAO, which is operated by AURA under cooperative
agreement with NSF.



{\it Facilities:} \facility{Sloan, Spitzer, XMM, Mayall}.

\appendix
 
\section{X-ray Observations}

We have reduced data from {\em XMM-Newton} on 2 fields in the XFLS
area, which is contained in this catalog (Table~\ref{tab:tab3}).  As
the programs that these fields were part of were not completed, the
data has not appeared elsewhere, but they are nevertheless useful and
we catalog them in Table~\ref{tab:xmm}.  Five of the 24 detected
sources are quasar candidates in our catalog and 12 of the sources
have known spectroscopic redshifts.  We have used the XAssist package
\citep{pg03} to reduce the data from these observations.  As this is a
fully automated processing routine, it is possible that more accurate
results could be obtained with more careful data reduction; however,
the XAssist results are more than suitable for our purposes here given
the incomplete nature of the observations.  In addition to the
coordinates and exposure times (ks), Table~\ref{tab:xmm} gives the
total flux, soft- and hard-band counts (background corrected),
hardness ratio (h-s)/(h+s) and its error.  We also denote which X-ray
sources appear in our quasar catalog (by the catalog ID number) along
with any previous identifications (see the references in
\S~\ref{sec:cat}, with two additional objects matched to
\citealt{lacy05} and \citealt{fms+06}).


\clearpage

\clearpage

\clearpage

\clearpage

\clearpage

\clearpage

\clearpage

\clearpage

\begin{deluxetable}{lcccccccc}
\tablewidth{0pt}
\tablecaption{Wide-area MIR Field Parameters\label{tab:tab0}}
\tablehead{
\colhead{Field} &
\colhead{RA} &
\colhead{Dec} &
\colhead{Area} &
\colhead{Exp} &
\colhead{$3.6$/$8.0\mu$m $5\sigma$ Depth} &
\colhead{$3.6$/$8.0\mu$m $95\%$ Comp. Depth} \\
\colhead{} &
\colhead{(deg)} &
\colhead{(deg)} &
\colhead{(deg$^2$)} &
\colhead{(s)} &
\colhead{($\mu$Jy)} &
\colhead{($\mu$Jy)}
}
\startdata
XFLS       & 259.5  & 59.5   & 4    & 60   & $\cdots$/$\cdots$ & 20(77\%)/100(94\%) \\
Bo\"{o}tes & 218.02 & 34.28  & 8.5  & 90   & 6.4/56    & $\cdots$/$\cdots$ \\
ELAIS-N1   & 242.75 & 55.0   & 9.3  & 120  & 3.7/37.8 & 14/56 \\
ELAIS-N2   & 249.2  & 41.029 & 4.2  & 120  & 3.7/37.8 & 14/56 \\
Lockman    & 161.25 & 58.0   & 11.1 & 120  & 3.7/37.8 & 14/56 \\
COSMOS     & 150.62 & 2.21   & 2.0  & 1200 & 0.9/14.6 & $\cdots$/$\cdots$ \\
\enddata
\end{deluxetable}

\begin{deluxetable}{llllll}
\tablewidth{0pt}
\tablecaption{Mean Observed and Theoretical Star and Quasar MIR Colors\label{tab:tab1}}
\tablehead{
\colhead{Color} &
\colhead{Star($\alpha=2$)} &
\colhead{Star(obs.)} &
\colhead{QSO($\alpha=-1$)} &
\colhead{QSO(obs.)} 
}
\startdata
$[3.6]-[4.5]$ & $-$0.485 & $-$0.497 & 0.242 & 0.287 \\
$[5.8]-[8.0]$ & $-$0.698 & $-$0.604 & 0.349 & 0.454 \\
$[3.6]-[8.0]$ & $-$1.734 & $-$1.534 & 0.867 & 1.162 \\
$[3.6]-[5.8]$ & $-$1.036 & $-$0.926 & 0.518 & 0.703 \\
$[4.5]-[8.0]$ & $-$1.249 & $-$1.031 & 0.625 & 0.853 \\
\enddata
\tablecomments{For $\alpha$ we adopt the nomenclature: $f_{\nu}\propto\nu^{\alpha}$.}
\end{deluxetable}


\begin{deluxetable}{lrrrr}
\tablewidth{0pt}
\tablecaption{MIR Color Selection Comparison\label{tab:tab2}}
\tablehead{
\colhead{} &
\colhead{All} &
\colhead{Point} &
\colhead{Extended} &
\colhead{Bright}
}
\startdata
N Objects & 52659 & 22473 & 30186 & 2225 \\
N 8-D & 5468 & 3426 & 2042 & 273 \\
N 6-D & 5222 & 3426 & 1796 & 271 \\
N Lacy & 15776 & 3424 & 12352 & 299 \\
N Stern & 5659 & 2981 & 2678 & 268 \\
N Ch1/2 Cut & 12360 & 3207 & 9153 & 474 \\
\enddata
\end{deluxetable}

\begin{deluxetable}{lcl}
\tabletypesize{\small}
\tablewidth{0pt}
\tablecaption{MIR/Optical Quasar Candidate Catalog Format\label{tab:cat}}
\tablehead{
\colhead{Column} &
\colhead{Format} &
\colhead{Description}
}
\startdata
1 & I7 & Unique catalog number \\
2 & F10.6 & Right ascension in decimal degrees (J2000.0) \\
3 & F10.6 & Declination in decimal degrees (J2000.0) \\
4 & A18 & Name: SDSS J$hhmmss.ss+ddmmss.s$ (J2000.0) \\
5 & A19 & SDSS Object ID \\
6 & F6.3 & $u$ PSF asinh magnitude (corrected for Galactic extinction) \\
7 & F6.3 & $g$ PSF asinh magnitude (corrected for Galactic extinction) \\
8 & F6.3 & $r$ PSF asinh magnitude (corrected for Galactic extinction) \\
9 & F6.3 & $i$ PSF asinh magnitude (corrected for Galactic extinction) \\
10 & F6.3 & $z$ PSF asinh magnitude (corrected for Galactic extinction) \\
11 & F8.3 & {\em Spitzer}-IRAC Channel 1 $3.6\mu$m flux density ($\mu$Jy) \\
12 & F8.3 & {\em Spitzer}-IRAC Channel 2 $4.5\mu$m flux density ($\mu$Jy) \\
13 & F8.3 & {\em Spitzer}-IRAC Channel 3 $5.8\mu$m flux density ($\mu$Jy) \\
14 & F8.3 & {\em Spitzer}-IRAC Channel 4 $8.0\mu$m flux density ($\mu$Jy) \\
15 & F6.3 & Error in PSF $u$ asinh magnitude \\
16 & F5.3 & Error in PSF $g$ asinh magnitude \\
17 & F5.3 & Error in PSF $r$ asinh magnitude \\
18 & F5.3 & Error in PSF $i$ asinh magnitude \\
19 & F5.3 & Error in PSF $z$ asinh magnitude \\
20 & F6.3 & Error in $3.6\mu$m flux density ($\mu$Jy) \\
21 & F6.3 & Error in $4.5\mu$m flux density ($\mu$Jy) \\
22 & F6.3 & Error in $5.8\mu$m flux density ($\mu$Jy) \\
23 & F6.3 & Error in $8.0\mu$m flux density ($\mu$Jy) \\
24 & F6.3 & $u$-band Galactic extinction, $A_u$ (mag); $A_u/A_g/A_r/A_i/A_z=5.155/3.793/2.751/2.086/1.479 \times E(B-V)$ \\
25 & I1 & SDSS Morphology (point$=6$, extended$=3$) \\
26 & I1 & Lacy wedge flag (in$=1$, out$=0$) \\
27 & I1 & Stern wedge flag (in$=1$, out$=0$) \\
28 & I1 & Modified Stern wedge flag (in$=1$, out$=0$) \\
29 & I1 & 8-D Bayesian classification (in$=1$, out$=0$) \\
30 & I1 & 6-D Bayesian classification (in$=1$, out$=0$) \\
31 & F6.3 & 8-D Photometric redshift (see \citealt{wrs+04})\\
32 & F6.3 & Lower limit of 8-D photometric redshift range \\
33 & F6.3 & Upper limit of 8-D photometric redshift range \\
34 & F6.3 & 8-D Photometric redshift range probability \\
35 & F6.3 & 6-D Photometric redshift (see \citealt{wrs+04})\\
36 & F6.3 & Lower limit of 6-D photometric redshift range \\
37 & F6.3 & Upper limit of 6-D photometric redshift range \\
38 & F6.3 & 6-D Photometric redshift range probability \\
39 & F5.3 & Optical selection flag (from Richards et al. 2008) \\
40 & F5.3 & Previous catalog identification \\
41 & F5.3 & Previous catalog object redshift
\enddata
\end{deluxetable}


\begin{deluxetable}{lllrlllllllll}
\tabletypesize{\scriptsize}
\rotate
\tablewidth{0pt}
\tablecaption{Bayesian Quasar Candidates\label{tab:tab3}}
\tablehead{
\colhead{} &
\colhead{R.A.} &
\colhead{Decl.} &
\colhead{Name} &
\colhead{} &
\colhead{} &
\colhead{} &
\colhead{} &
\colhead{} &
\colhead{} &
\colhead{$f_{3.6}$} &
\colhead{$f_{4.5}$} \\
\colhead{Number} &
\colhead{(deg)} &
\colhead{(deg)} &
\colhead{(SDSS J)} &
\colhead{ObjID} &
\colhead{$u$} &
\colhead{$g$} &
\colhead{$r$} &
\colhead{$i$} &
\colhead{$z$} &
\colhead{($\mu$Jy)} &
\colhead{($\mu$Jy)} \\
\colhead{(1)} &
\colhead{(2)} &
\colhead{(3)} &
\colhead{(4)} &
\colhead{(5)} &
\colhead{(6)} &
\colhead{(7)} &
\colhead{(8)} &
\colhead{(9)} &
\colhead{(10)} &
\colhead{(11)} &
\colhead{(12)}
}
\startdata
1\ldots & 149.326148 & 2.713954 & 095718.27+024250.2 & 587727944570503405 & 20.378 & 20.151
 & 19.805 & 19.581 & 19.705 & 126.810 & 185.392\\
2\ldots & 149.335600 & 2.488276 & 095720.54+022917.7 & 587726033308877340 & 22.961 & 22.082
 & 21.033 & 20.779 & 20.784 & 37.673 & 35.500\\
3\ldots & 149.349168 & 1.937556 & 095723.80+015615.2 & 587727943496761583 & 19.998 & 20.076
 & 19.893 & 19.914 & 19.977 & 61.660 & 103.770\\
5\ldots & 149.358575 & 2.750292 & 095726.05+024501.0 & 587727944570569013 & 20.661 & 20.946
 & 20.572 & 20.453 & 20.477 & 58.928 & 81.784\\
\enddata
\end{deluxetable}


\begin{deluxetable}{lllrlllllllll}
\tabletypesize{\scriptsize}
\rotate
\tablewidth{0pt}
\tablecaption{Wedge Quasar Candidates\label{tab:tab4}}
\tablehead{
\colhead{} &
\colhead{R.A.} &
\colhead{Decl.} &
\colhead{Name} &
\colhead{} &
\colhead{} &
\colhead{} &
\colhead{} &
\colhead{} &
\colhead{} &
\colhead{$f_{3.6}$} &
\colhead{$f_{4.5}$} \\
\colhead{Number} &
\colhead{(deg)} &
\colhead{(deg)} &
\colhead{(SDSS J)} &
\colhead{ObjID} &
\colhead{$u$} &
\colhead{$g$} &
\colhead{$r$} &
\colhead{$i$} &
\colhead{$z$} &
\colhead{$\mu$Jy} &
\colhead{$\mu$Jy} \\
\colhead{(1)} &
\colhead{(2)} &
\colhead{(3)} &
\colhead{(4)} &
\colhead{(5)} &
\colhead{(6)} &
\colhead{(7)} &
\colhead{(8)} &
\colhead{(9)} &
\colhead{(10)} &
\colhead{(11)} &
\colhead{(12)}
}
\startdata
4\ldots & 149.351001 & 1.955768 & 095724.24+015720.7 & 587727943496762081 & 24.292 & 21.987
 & 20.286 & 19.699 & 19.323 & 146.039 & 131.027\\
17\ldots & 149.390413 & 2.919368 & 095733.69+025509.7 & 587726033845813857 & 23.405 & 23.41
5 & 22.363 & 21.831 & 22.037 & 137.255 & 125.932\\
26\ldots & 149.404266 & 1.737216 & 095737.02+014413.9 & 587726032235135314 & 22.585 & 21.36
8 & 20.113 & 19.658 & 19.406 & 110.340 & 100.392\\
30\ldots & 149.413157 & 1.668670 & 095739.15+014007.2 & 587726032235135709 & 24.339 & 22.69
9 & 21.685 & 21.067 & 20.674 & 42.732 & 42.716\\
\enddata
\end{deluxetable}

\begin{deluxetable}{lllrlrrlllll}
\tabletypesize{\scriptsize}
\rotate
\tablewidth{0pt}
\tablecaption{{\em XMM-Newton} Objects in the XFLS Area\label{tab:xmm}}
\tablehead{
\colhead{Name} &
\colhead{RA} &
\colhead{Dec} &
\colhead{Exp.} &
\colhead{Total Flux} &
\colhead{Soft Cnts} &
\colhead{Hard Cnts} &
\colhead{HR} &
\colhead{$\sigma_{\rm HR}$} &
\colhead{CATID} &
\colhead{Ref/ID} &
\colhead{z}
}
\startdata
X171007.10+591127.7 & 257.529588 & 59.191014 & 12.8 & 6.98477e-14 & 23.1 & 8.8 & -0.45 & 0.34 & & Lacy05 & \\
X171029.30+590833.7 & 257.622073 & 59.142688 & 12.8 & 2.53155e-13 & 86.2 & 27.0 & -0.52 & 0.12 & & DR5QSO & 0.864 \\
X171049.65+590802.9 & 257.706889 & 59.134134 & 12.8 & 8.59703e-14 & 43.6 & 11.0 & -0.60 & 0.21 & 5606 & P06QSO & 1.234 \\
X171126.68+585541.8 & 257.861152 & 58.928266 & 12.8 & 2.86859e-13 & 123.3 & 24.9 & -0.66 & 0.11 & 5629 & DR5QSO & 0.537 \\
X171136.73+590115.7 & 257.903044 & 59.021019 & 12.8 & 4.43253e-14 & 21.1 & 15.7 & -0.15 & 0.23 & & &  \\
X171156.98+591220.0 & 257.987412 & 59.205546 & 12.8 & 6.17736e-14 & 21.3 & 9.2 & -0.40 & 0.34 & 5648 & P06QSO & 2.043 \\
X171159.26+590433.1 & 257.996904 & 59.075852 & 12.8 & 4.83567e-14 & 18.8 & 11.4 & -0.24 & 0.25 & 5651 & &  \\
X171231.71+591217.6 & 258.132137 & 59.204879 & 12.8 & 1.13168e-13 & 22.0 & 10.2 & -0.37 & 0.29 & & &  \\
X171634.82+594310.9 & 259.145078 & 59.719695 & 9.3 & 1.15027e-13 & 15.6 & 21.1 & 0.15 & 0.57 & & &  \\
X171638.06+594514.6 & 259.158591 & 59.754067 & 11.5 & 5.98434e-14 & -0.7 & 23.4 & 1.00 & 1.77 & & &  \\
X171641.88+593758.7 & 259.174511 & 59.632980 & 9.3 & 2.5577e-13 & 39.4 & 20.2 & -0.32 & 0.37 & & &  \\
X171652.39+593543.6 & 259.218298 & 59.595450 & 9.3 & 3.81194e-14 & -5.1 & 13.1 & 1.00 & 2.36 & & &  \\
X171712.89+593828.7 & 259.303714 & 59.641318 & 11.5 & 2.6894e-13 & & & & & 5845 & P06QSO & 0.233 \\
X171717.25+594640.4 & 259.321867 & 59.777894 & 9.3 & 2.13858e-13 & 51.9 & 27.6 & -0.31 & 0.23 & & & \\
X171736.53+593010.9 & 259.402214 & 59.503022 & 11.5 & & & & & & & P96QSO & 0.599 \\
X171737.02+593011.1 & 259.404248 & 59.503075 & 8.4 & 1.73472e-13 & 35.0 & 8.6 & -0.61 & 0.35 & 5854 & DR5QSO & 0.599 \\
X171746.28+594123.7 & 259.442816 & 59.689926 & 1.1 & 3.24386e-13 & 40.0 & 18.4 & -0.37 & 0.25 &  & Fadda06 &  \\
X171747.39+593258.9 & 259.447471 & 59.549688 & 9.3 & 4.64486e-13 & 97.3 & 51.9 & -0.30 & 0.16 & 5858 & P06Sy1 & 0.248 \\
X171802.80+594001.0 & 259.511675 & 59.666935 & 11.5 & 1.59579e-13 & & & & & & &  \\
X171806.56+593312.6 & 259.527341 & 59.553487 & 9.3 & 8.9476e-13 & 227.2 & 51.9 & -0.63 & 0.10 & 5878 & P06QSO & 0.273 \\
X171839.19+593402.0 & 259.663301 & 59.567217 & 9.3 & 3.81391e-13 & 101.4 & 49.4 & -0.34 & 0.14 & 5899 & P06QSO & 0.383 \\
X171902.17+593715.7 & 259.759039 & 59.621019 & 9.3 & 9.07265e-13 & 306.9 & 92.8 & -0.54 & 0.08 & 5914 & DR5QSO & 0.179 \\
X171943.91+594100.0 & 259.932958 & 59.683335 & 9.3 & 5.13025e-13 & 27.1 & -6.2 & -1.00 & 1.50 & & P06QSO & 0.129
\enddata
\end{deluxetable}

\end{document}